\documentclass[]{polymathic}
\usepackage{amssymb}
\usepackage{multirow}
\usepackage{bigdelim}
\usepackage{todonotes}
\usepackage{longtable}
\usepackage{tabularray}
\usepackage{subcaption}
\usepackage{wrapfig}
\usepackage[most]{tcolorbox} 
\usepackage{xcolor}
\usepackage{url}
\usepackage{wrapfig}
\usepackage{graphicx}
\usepackage{booktabs}   
\usepackage{tabularx}   
\usepackage{caption}    
\usepackage[sort&compress]{natbib}  
\usepackage{aas_macros}

\title{AION-1: Omnimodal Foundation Model for Astronomical Sciences}

\author[*,1,2,3,4]{Liam Parker}
\author[*,5,2]{Francois Lanusse}
\author[*,6]{Jeff Shen}
\author[7]{Ollie Liu}
\author[8]{Tom Hehir}
\author[3]{Leopoldo Sarra}
\author[3]{Lucas Meyer}
\author[9]{Micah Bowles}
\author[2,3]{Sebastian Wagner-Carena}
\author[2]{Helen Qu}
\author[2,3]{Siavash Golkar}
\author[2]{Alberto Bietti}
\author[10]{Hatim Bourfoune}
\author[10]{Nathan Cassereau}
\author[10]{Pierre Cornette}
\author[2,11]{Keiya Hirashima}
\author[2]{Geraud Krawezik}
\author[2]{Ruben Ohana}
\author[3]{Nicholas Lourie}
\author[2,3]{Michael McCabe}
\author[2]{Rudy Morel}
\author[1,8]{Payel Mukhopadhyay}
\author[12]{Mariel Pettee}
\author[2]{Bruno Regaldo-Saint Blancard}
\author[3]{Kyunghyun Cho}
\author[8]{Miles Cranmer}
\author[2,3,6]{Shirley Ho}

\affiliation[1]{University of California, Berkeley}
\affiliation[2]{Flatiron Institute}
\affiliation[3]{New York University}
\affiliation[4]{Lawrence Berkeley National Laboratory}
\affiliation[5]{Université Paris-Saclay,
Université Paris Cité, CEA, CNRS, AIM}
\affiliation[6]{Princeton University}
\affiliation[7]{University of Southern California}
\affiliation[8]{University of Cambridge}
\affiliation[9]{University of Oxford}
\affiliation[10]{IDRIS, CNRS}
\affiliation[11]{RIKEN Center for iTHEMS}
\affiliation[12]{University of Wisconsin--Madison}

\contribution[*]{Equal Contribution.}

\abstract{
While foundation models have shown promise across a variety of fields, astronomy still lacks a unified framework for joint modeling across its highly diverse data modalities. In this paper, we present AION-1, a family of large-scale multimodal foundation models for astronomy. AION-1 integrates heterogeneous imaging, spectroscopic, and scalar data using a two-stage architecture: modality-specific tokenization followed by transformer-based masked modeling of cross-modal token sequences. The model is pretrained on five large-scale surveys: Legacy Survey, Hyper Suprime-Cam (HSC), Sloan Digital Sky Survey (SDSS), Dark Energy Spectroscopic Instrument (DESI), and Gaia. These span more than 200 million observations of stars, galaxies, and quasars. With a single frozen encoder, AION-1 achieves strong results on a broad suite of downstream tasks, including galaxy and stellar property estimation, galaxy morphology classification, similarity-based retrieval, galaxy image segmentation, and spectral super-resolution. We release AION-1 model variants ranging from 300 M to 3.1 B parameters. Beyond astronomy, AION-1 provides a scalable blueprint for multimodal scientific foundation models that can seamlessly integrate noisy, instrument-specific observations. All code, tokenizers, pretrained weights, and a lightweight evaluation suite are released under an open-source license.}

\date{\today}
\titlecorrespondence{Liam Parker: \email{lhparker@berkeley.edu}; Francois Lanusse: \email{francois.lanusse@cnrs.fr}}
\titlecode{\url{https://github.com/PolymathicAI/AION/}}

\begin{document}

\maketitle

\section{Introduction}
\label{sec:intro}

\begin{figure*}
    \centering
    \includegraphics[width=\textwidth]{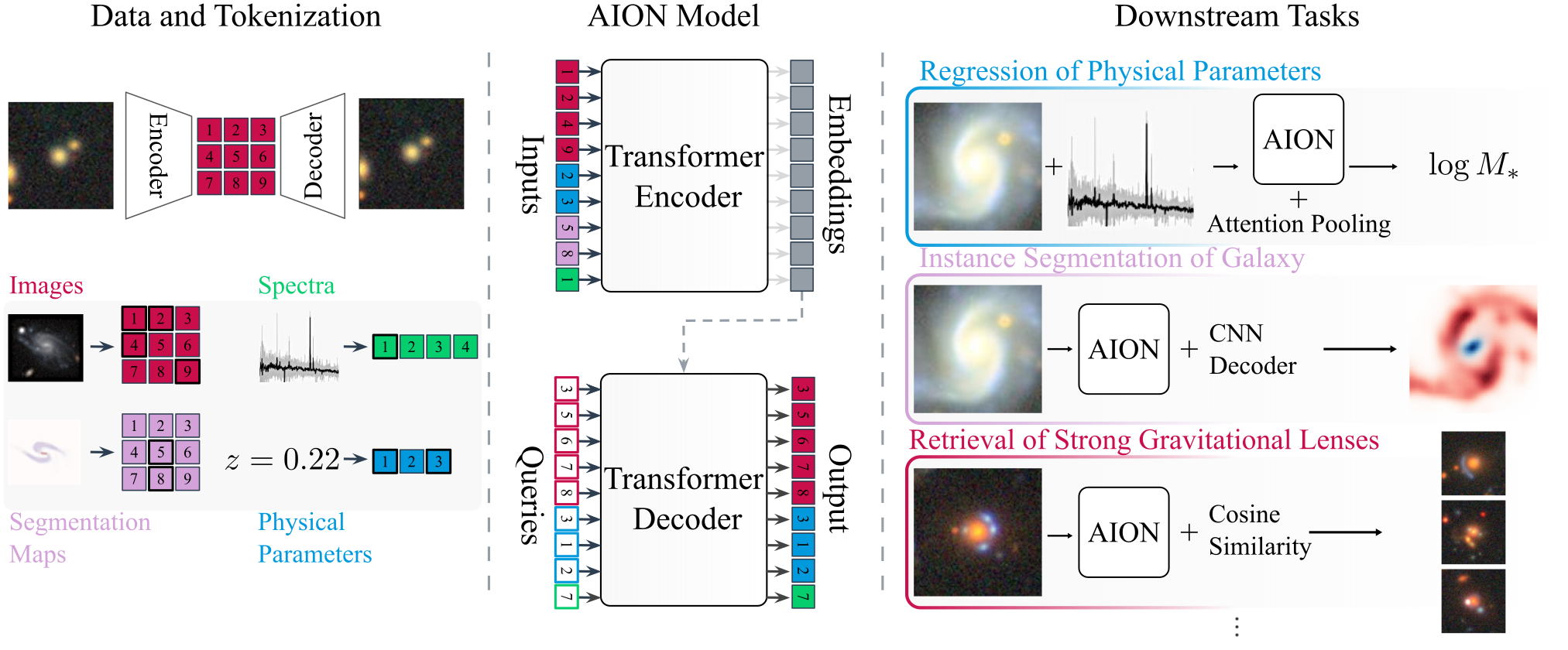}
    \caption{AION-1 integrates 39 different data modalities — multiband images, optical spectra, and various properties and measurements — into a single model usable for a wide range of downstream applications. It implements a two-step process: first, bespoke tokenization strategies that homogenize the diverse scientific data, followed by multimodal masked modeling that learns how different observations relate, inducing a deep understanding of the underlying physical objects. Astronomers can then leverage AION-1’s rich astrophysical understanding for a variety of downstream tasks.}
    \label{fig:main-figure}
\end{figure*}

Foundation models have transformed natural language processing and computer vision \citep{achiam2023gpt, team2024gemini, dubey2024llama}. However, this class of models has not been fully explored in scientific domains where data are often complex and heterogeneous, combining multiple instruments, measurement protocols, and noise sources unique to real-world experiments. As a result, many scientific analyses employ bespoke models that treat each modality in isolation or rely on strict - often hand-crafted - schemas for cross-modal data fusion. 

Within the broader scientific landscape, astronomy provides a particularly compelling testbed for the development of multimodal scientific foundation models owing to both the volume of publicly available data and its extraordinary diversity of measurements. Indeed, recent works have begun to explore multi-modal foundation models in astronomy \citep{Parker2024, Mishra-Sharma2024, Zhang2024, rizhko2024astrom3}; however, these approaches have been limited to single physical phenomena and relied primarily on contrastive objectives, which face fundamental limitations including generalization to arbitrary modalities and difficulty in capturing information beyond the mutual information between modalities. 

In this paper, we introduce AION-1 (AstronomIcal Omni-modal Network), a large-scale multimodal foundation model for astronomy designed to handle \emph{arbitrary} numbers of modalities across multiple physical phenomena. AION-1 unifies imaging, spectroscopy, photometry, and other object-level measurements from major ground- and space-based observatories into a single model for galaxies, stars, and quasars. By bridging these disparate data types, AION-1 addresses a key challenge in scientific machine learning: the integration of multiple heterogeneous datasets spanning different instruments, measurement protocols, noise sources, and physical phenomena into a single, unified framework.

At the heart of AION-1 lies a two-step approach: \textbf{Universal Tokenization of Diverse Data}, where we homogenize real-world scientific observations with discrete quantization across different data types, instruments, and observatories, followed by \textbf{Multimodal Masked Modeling}, where we train a single transformer encoder-decoder with a masked-token objective over all modalities simultaneously. 
Once trained, we demonstrate \emph{emergent behaviors} in the AION-1 models that reflect the potential for multimodal scientific foundation models to capture non-trivial physical insights from raw data alone:
\begin{itemize}
\item \textbf{Emergent Physical Understanding.} AION-1 can solve non-trivial scientific tasks using only a simple linear head on top of its learned representations. 
\item \textbf{Superior Performance in the Low-Data Regime.} AION-1 can achieve competitive results on downstream inference tasks even with orders of magnitude less data than its supervised counterparts. 
\item \textbf{Flexible Data Fusion.} AION-1 can use arbitrary combinations of observations, enabling seamless data fusion on downstream tasks as well as cross-modal conditional generation.
\item \textbf{Physical Structure of the Latent Space}: AION-1’s embedding space organizes objects along physically meaningful directions, enabling powerful retrieval of rare observations that surpasses current state-of-the-art retrieval methods in astronomy. 
\end{itemize}

Beyond astronomy, the data tokenization strategies, masked modeling, and cross-modal generation strategies introduced address key challenges in real-world scientific data—namely, heterogeneity, noise, and instrument-specific idiosyncrasies. Moreover, by focusing on purely observational data, our approach is applicable in any data-rich field, even when strong physical models are not available. 
\subsection{Contributions}
In summary, we present the following contributions:
\begin{itemize}
\item  We present AION-1, a family of token-based multimodal scientific foundation models ranging in size from 300M to 3.1B parameters. AION-1 is a large-scale model designed for arbitrary combinations of highly heterogeneous scientific observations.
\item We develop bespoke tokenization methods to homogenize a wide variety of astronomical data into a single coherent corpus. These innovations address the heterogeneity, noise, and instrument-specific peculiarities that challenge standard scientific modeling.
\item We demonstrate that AION-1 achieves competitive to state-of-the-art performance on a broad range of scientific tasks with even simple probing, while significantly outperforming supervised baselines in low-data regimes, rendering the model highly usable by downstream researchers even without dedicated finetuning.
\end{itemize}
By tackling the challenges of data heterogeneity, noise, and diverse instrumentation, AION-1 offers a promising paradigm for future multimodal foundation models beyond astronomy, setting the stage for a new era of large-scale, cross-domain scientific exploration.

\section{Related Work}

Multimodal foundation models have become a cornerstone of modern self-supervised learning \citep{achiam2023gpt, team2024gemini, dubey2024llama, liu2023llava, claude, stabilityai2022stable}. Indeed, recent advances like GPT-4V \citep{achiam2023gpt}, Claude 3 \citep{claude}, and LLaVA \citep{liu2023llava} have achieved human-level performance in visual reasoning, while models like Imagen \citep{saharia2022imagen} and Stable Diffusion \citep{stabilityai2022stable} have enabled high-quality image generation from text.
However, these models primarily rely on language to bridge modalities, which is often unavailable for scientific data. Recent work on early-fusion models, such as Chameleon \citep{chameleonteam}, 4M \citep{mizrahi20234m}, or PercieverIO \citep{perceiverio}, have demonstrated promising alternatives by learning mappings between modalities. 

While these methodological advances in foundation models have transformed many fields, astronomy presents unique challenges, including heterogenous instruments, measurement protocols, and noise. As such, astronomy-specific efforts have emerged. For example, supervised pre-trained models like Zoobot \citep{Walmsley2024} have leveraged 100M human annotations for galaxy morphology prediction obtained through extensive citizen science campaigns. Large-scale, self-supervised approaches trained on single-modal data have also emerged, including transformer-based models for Gaia stellar data \citep{Leung2024}, APOGEE spectra \citep{Koblischke2024} and astronomical images \citep{Smith2024} and contrastive approaches for astronomical images \citep{Hayat2021, stein2021self, stein2022mining}. Finally, recent multimodal contrastive approaches have been introduced, starting with galaxy image-spectra pairs in \citet{Parker2024} and followed by galaxy images and text \citep{Mishra-Sharma2024} and time-series and photometry \citep{Zhang2024, rizhko2024astrom3}.

Relative to these methods, AION-1 represents an advance in both scale and scope: we train a multimodal model to billion-parameter scales and attempt to unify arbitrary modalites or different object types; in this case, 39 modalities across 200 million unique measurements spanning galaxies, stars, and quasars. 

\section{Data}

\begin{figure}[ht]
  \centering
  \begin{subfigure}[ht]{0.4\textwidth}
    \centering
    \includegraphics[width=0.9\linewidth]{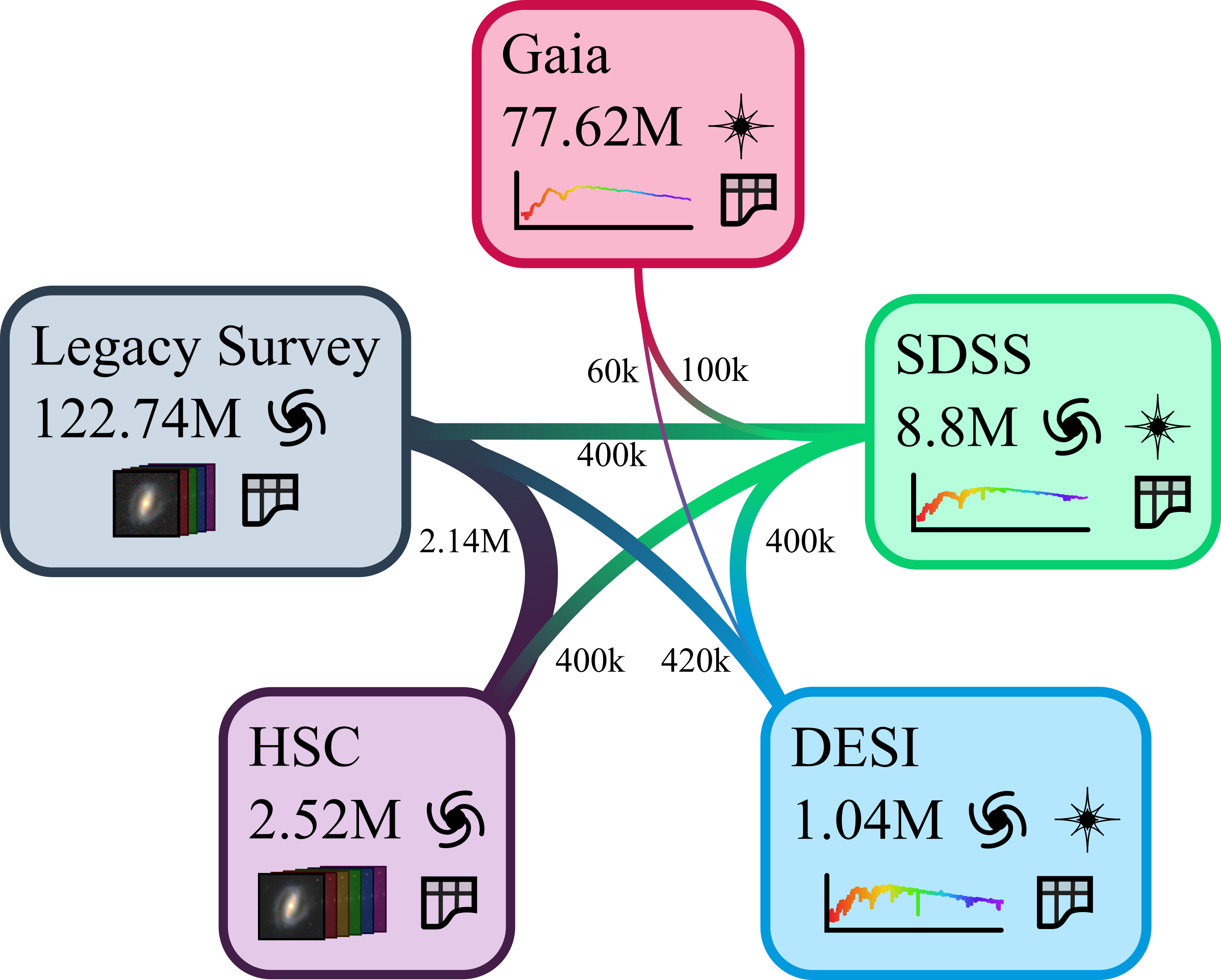}
  \end{subfigure}\hfill
  \begin{subtable}[ht]{0.58\textwidth}
    \centering
    \small
    \begin{tabularx}{\linewidth}{@{}lX@{}}
      \toprule
      \textbf{Survey} & \textbf{Modalities} \\ \midrule
      Legacy Survey &
        4-band images; $g,r,i,z$ fluxes; WISE $W1$–$W4$ fluxes; $E(B-V)$ extinction;
        ellipticity $(e_{1},e_{2})$; half-light radius $R_{\rm eff}$ \\[3pt]
      HSC &
        5-band images; $g,r,i,z,y$ fluxes and extinction; shape catalog
        $(\gamma_{11},\gamma_{12},\gamma_{22})$ \\[3pt]
      SDSS & Optical spectrum; redshift ($z$) \\[3pt]
      DESI & Optical spectrum; redshift ($z$) \\[3pt]
      Gaia &
        BP/RP spectra; parallax $\varpi$; sky coords $(\alpha,\delta)$;
        fluxes $G$, BP, RP \\ \bottomrule
    \end{tabularx}
  \end{subtable}
  \caption{\textbf{AION-1 pre-training inputs.} \textit{Left}: visual representation of the various surveys used during pretraining, their total volume (in number of objects), and the size of the cross-matches between datasets. \textit{Right}: A more detailed break-down of the modalities provided by each survey.}
  \label{fig:mmu-components}
\end{figure}

AION-1 is pretrained on the publicly available data from \cite{MMU} (hereafter MMU), a large-scale dataset of ML-ready, multimodal astronomical data. We use five surveys: Hyper Suprime-Camera (HSC) \citep{aihara2018hyper} and Legacy Imaging Survey \citep{dey2019overview} for galaxy images; DESI \citep{aghamousa2016desi} and SDSS \citep{york2000sloan} for high-resolution spectra and cosmic distances on galaxies and stars; and Gaia \citep{gaia2016gaia} for low-resolution spectra and precise photometric and astrometric measurements for stars in the Milky Way. The relative contribution and modalities present in each survey, along with the size of the cross-matches, is illustrated in \autoref{fig:mmu-components}. 

AION-1's pretraining emphasizes learning relationships between diverse observations of the same astronomical objects. Unlike 4M \citep{mizrahi20234m}, which requires all modalities simultaneously, we use pairwise associations across surveys. This flexibility accommodates uneven associations, and even enables the model to understand relationships between unpaired measurements, demonstrating transfer behavior. Below, we provide a brief overview of the details and data types from each survey, but refer the reader to \cite{MMU} or the original source dataset paper for more exhaustive descriptions. We note that the per-survey magnitude/quality thresholds, footprint choices, Gaia XP availability, and our reciprocal cross-match together define the effective selection function of the pre-training corpus; this shapes which morphologies, redshifts, environments, and S/N regimes might be emphasized in AION-1’s embeddings (see \cref{subsec:limitations}).

\subsection{Legacy Surveys}
\label{sec:data:ls}
The DESI Legacy Imaging Surveys combine three wide-field programs on the Blanco, Mayall, and Bok telescopes, delivering uniform \mbox{$\{g,r,i,z\}$} imaging over $\sim$20\,000deg$^{2}$—roughly half the sky. The galaxies are imaged at a pixel scale of 0.262 arcsec. MMU provides $160\!\times\!160$ pixel postage stamp cut-outs centered on galaxies from the Data Release 10 \citep{dey2019overview}, which we crop to $96\!\times\!96$ images; we use only objects in the Southern Galactic Cap, and retain objects with $\texttt{mag\_z} < 21$ that pass the MMU quality cuts, corresponding to roughly 122 million galaxies. 

\textbf{Modalities} Each object is packaged as: 
(1) calibrated \mbox{$\{g,r,i,z\}$} integrated fluxes and their inverse-variance estimates,  
(2) mid-IR fluxes W1–W4 from \emph{WISE},  
(3) Milky-Way reddening $E(B{-}V)$, and  
(4) basic shape/size descriptors—the ellipticity components $(e_{1},e_{2})$ and circularised half-light radius $R_{\rm eff}$—derived from the Legacy \texttt{tractor} model fits.

\subsection{Hyper Suprime-Cam (HSC)}
\label{sec:data:hsc}
The Hyper Suprime–Cam Subaru Strategic Program delivers deep, high‑resolution $\{g,r,i,z,y\}$ imaging over $\sim$1\,200\,deg$^{2}$. We use only the \textbf{wide} subset from PDR3 \citep{aihara2018hyper}. From the co–added \texttt{calexp} frames, MMU extracts $160\!\times\!160$‑pixel cut-outs at a pixel scale of 0.162 arcsec centered on catalog sources, which we crop to $96 \times 96$, as with the Legacy Survey images. Objects are kept when they satisfy: \texttt{mag\_i}~$<22.5$; at least three visits in every band (``full‑depth full‑colour''); and the standard HSC quality flags remove bright‑star contamination, edge artefacts, saturation and unreliable \texttt{cmodel} photometry. The resulting sample contains roughly 2.5 million galaxies.

\textbf{Modalities} Each object is packaged as: 
(1) calibrated \mbox{$\{g,r,i,z,y\}$} integrated fluxes and their inverse-variance estimates, 
(2) PSF-homogenised forced photometry in each band with extinction corrections, and  
(3) the moment-based SDSS shape tensor components $(\gamma_{11},\gamma_{12},\gamma_{22})$ computed by the HSC pipeline.

\subsection{Sloan Digital Sky Survey (SDSS)}
\label{sec:data:sdss}
The Sloan Digital Sky Survey (SDSS) \citep{ahumada202016th} has obtained medium‑resolution ($R\!\sim\!2\,000$) optical spectra for millions of objects. We use the aggregated public optical spectra from the Legacy, SEGUE‑1/2, BOSS and eBOSS programs\footnote{We note that the SDSS and BOSS instruments have different fiber aperture sizes, but in the present work we include them in the same dataset.}, covering 3\,650–10\,400\AA\, with resolutions of $R = \lambda /\Delta \lambda = 1,500$ at $3,800$\AA\ and $R=9,000$ at 9,000\AA. We keep only primary, science–target spectra from plates flagged \texttt{PLATEQUALITY=\textquotesingle good\textquotesingle}. This yields $\sim$4\,million galaxies and stars.

\textbf{Modalities.}  Each object is packaged as: (1) the optical spectrum, its inverse-variance estimates, and its wavelength and (2) the pipeline redshift.

\subsection{Dark Energy Spectroscopic Instrument (DESI)}
\label{sec:data:desi}
The Dark Energy Spectroscopic Instrument \citep{desi_survey} survey is collecting spectra for $\sim$40 million galaxies and quasars; MMU presently ingests the Early Data Release (EDR, 1\% of the full survey) \citep{desi_edr}.  Each spectrum spans 3\,600–9\,800\AA\ on a fixed 7,081‑pixel grid at resolutions of $R=2000$ at $3,600$\AA\ and $R=5,500$ at $9,000$\AA\, and is distributed with flux, wavelength and inverse‑variance arrays.  We select spectra from the SV3 ``one‑percent'' survey where \texttt{SV\_PRIMARY} is true, \texttt{OBJTYPE='TGT'} and \texttt{COADD\_FIBERSTATUS=0}, giving roughly 1 million galaxies, stars, and quasars.

\textbf{Modalities.}  Each object is packaged as: (1) the optical spectrum, its inverse-variance estimates, and its wavelength and (2) the pipeline redshift.

\subsection{Gaia}
\label{sec:data:gaia}
Gaia DR3 \citep{gaia2016gaia} provides low‑resolution prism spectra from its blue (BP) and red (RP) photometers for 220 million Milky‑Way sources in addition to precise astrometry and broad‑band photometry. MMU stores each BP/RP spectrum as the 110 Gauss–Hermite coefficients released by the mission (55 BP + 55 RP), which can be resampled onto an 1\,101‑pixel wavelength grid via \texttt{GaiaXPy}.  We include all DR3 objects that have a mean BP/RP spectrum, retaining the full set of associated photometric, astrometric and stellar‑parameter metadata. 

\textbf{Modalities.}  Each object is packaged as:
(1) The 110 BP/RP spectral coefficients, 
(2) four-parameter astrometry (sky coordinates and parallax), and (3) mean fluxes in the $G$\footnote{$G$ is the mission’s very broad “white-light” band measured by the astrometric field CCDs.}, BP and RP bands.

\subsection{Cross-matching strategy}  
For each pair of surveys we perform a nearest-neighbour match within a $1$ arcsec radius on the sky and keep only reciprocal matches. Every resulting match is materialised as its own dataset. Each object in these datasets therefore aggregates all modalities from both parent surveys so that a single file read yields a fully fused, multi-survey view of the same astrophysical object. During AION-1 pre-training we draw samples both from the individual survey datasets and from these cross-matched sets, as detailed in the next sections. We note that this procedure may preferentially retain bright, isolated, well-centered sources and may de-emphasize blended or offset systems, introducing a further selection effect on the joint training distribution (see \cref{subsec:limitations}).

\section{Tokenization of Astronomical Data Modalities}

Tokenization in AION-1 transforms heterogeneous data into a unified, transformer-compatible representation. Astronomical datasets present two key challenges: the variety of data types (2D images, 1D spectra, scalar values) and the diversity of sources within each type (different telescopes, resolutions, and instrument formats). We address this through modality-specific tokenizers that provide intra-modality standardization; each modality uses a dedicated tokenizer capable of handling multiple instruments, ensuring aligned representations within each data type. Moreover, the need to train multiple tokenizers for a modality with multiple survey inputs is removed.

\subsection{Multi‑Survey Image Tokenizer}
\label{sec:image_tokenizer}

\subsubsection{Preprocessing} 

Each input from an imaging survey provides (i) a per-band flux map $\mathbf{x}$, (ii) a pixel-wise inverse variance map $\mathbf{\Sigma}$, and (iii) a per-pixel mask $\mathbf{m}$ for a given source. Our tokenizer ingests heterogeneous measurements drawn from both HSC (five filters $\{g,r,i,z,y\}$) and the Legacy Survey SGC (LS; four filters $\{g,r,i,z\}$). The two pipelines vary in central wavelength, pixel scale, zero‑point, and noise. Therefore, we treat all bands from the surveys as distinct from eachother; i.e. $g$ from Legacy Survey is treated as a different band than $g$ from HSC. We stack all distinct bands into a single \emph{fixed} set of 9 channels (5 from HSC and 4 from LS), assigning a specific index to each channel. Next, we map every image into a 9‐channel tensor, filling the subset of channels corresponding to that image’s bands with flux values and setting any unused channels to zero; a binary mask $\mathbf m_c\in\{0,1\}^C$ tracks which bands are zeroed out. The result of this process is that all images drawn have the same dimension, and can be stacked into a single, heterogenous batch, while maintaining survey-specific provenance information. We then normalize the zero-points between surveys by rescaling HSC to the Legacy Survey zero-point of $22.5$\,mag via \(s = 10^{(\mathrm{ZP}-22.5)/2.5}\!\), and multiply by the ratio of pixel scales. While these steps are not strictly necessary - as the bands are already separated above - we find that it helps with training stability. Finally, we apply an $\mathrm{arcsinh}$ normalization to the images to account for their high dynamic range, which we invert before computing the autoencoding loss. We find that adequate range compression is crucial for training stability.

\subsubsection{Architecture and quantization.}

\paragraph{Subsampled Linear Projection}
Given a batch of images, \(\mathbf{x}\!\in\!\mathbb{R}^{B\times C\times H\times W}\) (batch by channels by height by width), we project it to a higher-dimensional space of size \(\mathrm{dim}_\mathrm{out}\approx 6C\) using  
\[
    \hat{\mathbf{x}} \;=\; \alpha(\mathbf{m}_c)\bigl(\tilde{\mathbf{x}}W + b\bigr),
\]
with learnable \(W\!\in\!\mathbb{R}^{C\times\mathrm{dim}_\mathrm{out}}\) and \(b\!\in\!\mathbb{R}^{\mathrm{dim}_\mathrm{out}}\). The scale factor \(\alpha(\mathbf{m})\) keeps the feature norm invariant to missing channels. The projection is inverted after decoding. We introduce the subsampled linear projection to expand each image into a higher-dimensional embedding that disentangles survey-specific channel information while preserving feature norms even when some bands are missing. 

\paragraph{Autoencoder}
Once subsampled, we feed the output of the subsampled linear projection, which is now a $54$-dimensional image, into a ResNet‐based autoencoder. Specifically, we use the MagViT architecture adapted from \cite{yu2023magvit}, in which we remove transformer blocks. The encoder therefore consists of 2 downsampling ResNet blocks, which reduce the dimensionality of the input image by a factor of 16, resulting in a laten space that is $24 \times 24 \times 512$; this is compressed to $d=4$ dimensions before being fed to the quantizer. The output of the quantizer is then projected back to $512$ dimensions, before being upsampled in the decoder. In total, the ResNet-based autoencoder has roughly 50M learnable parameters. 

\begin{wrapfigure}{r}{0.5\textwidth}
  \vspace{-4em}
  \centering
  \includegraphics[width=0.49\textwidth]{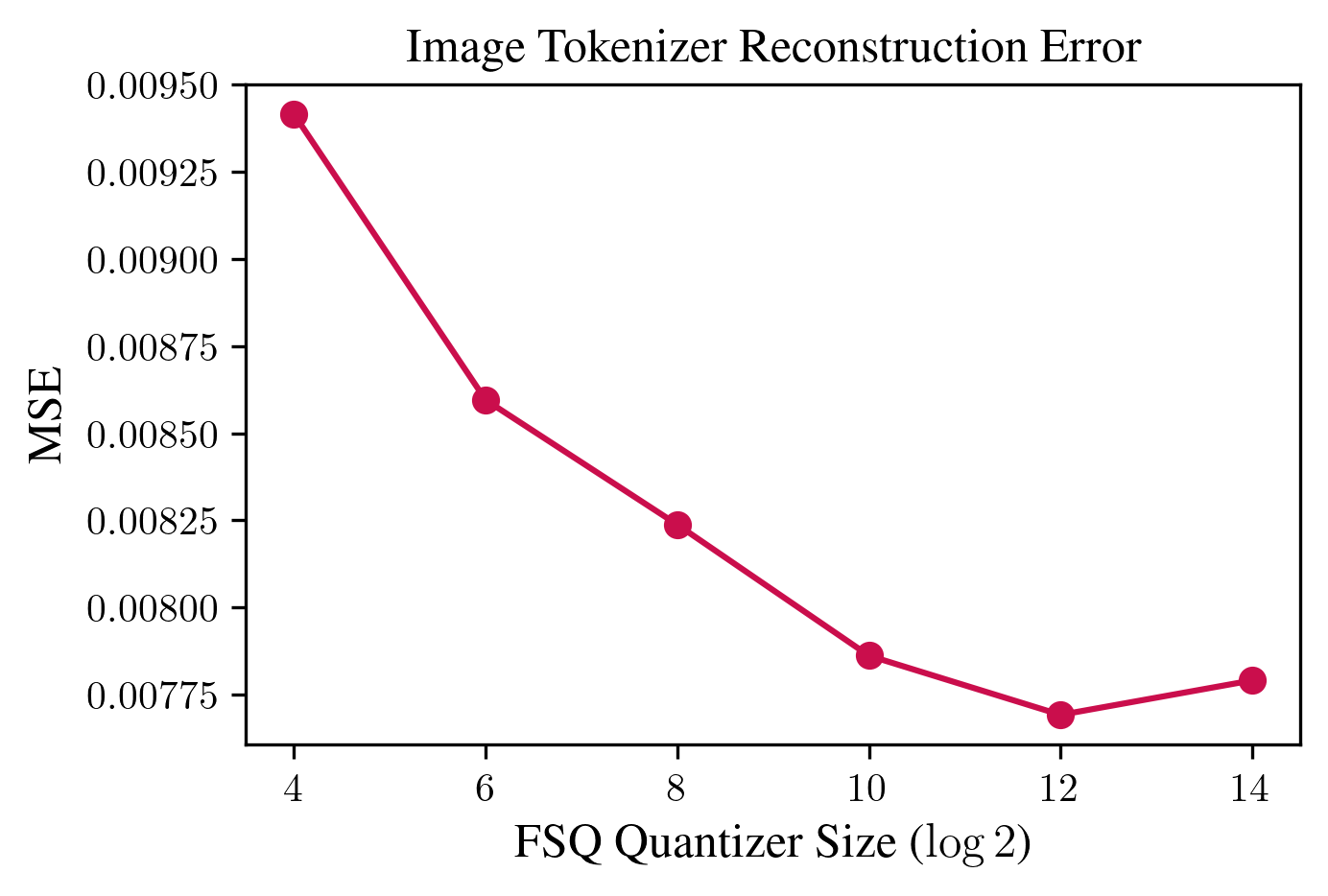}
  \caption{\textbf{Tokenizer Codebook Scaling}: We present the image tokenizer reconstruction error (MSE) as a function of FSQ quantizer size. We choose \(2^{12}\) as our ultimate codebook size for images, as it reconstruction loss appears to plateau around this point.}
  \label{fig:tokenizer_mse_scaling}
\end{wrapfigure}

\paragraph{Quantizer}
At the bottleneck of the tokenizer, we quantize features into a discrete set of codes. We experiment with multiple approaches, but empirically, we find that Finite Scale Quantization \citep[FSQ;][]{mentzer2023finite} yields the best performance in terms of reconstruction fidelity and training stability. Further, to explore the trade‐off between reconstruction loss and codebook utilization, we vary the codebook size from smaller (e.g., \(2^4\)) to larger (e.g., \(2^{14}\)) - following the recommended configurations in the FSQ paper - and observe that a size of \(2^{12}\) offers a desirable balance: the reconstruction loss plateaus with larger codebooks, while code usage remains sufficiently high to avoid underfitting with smaller codebooks; see \autoref{fig:tokenizer_mse_scaling} for reference. Consequently, our final configuration employs FSQ with codebook levels of $n_i=\{8,5,5,5\}$, equating to a rough size of $2^{12}$ codes. 

\subsubsection{Loss Function and Per‐band Weighting}
The tokenizer is trained using an inverse-variance-weighted Gaussian negative log-likelihood (NLL) that leverages our prior knowledge of the noise properties in each image, as reported by the data-generation pipelines. The NLL is given by:
\begin{equation}
   \mathcal{L}_\mathrm{NLL} \;=\; \sum_i \frac{1}{2} \parallel \mathbf{\Sigma_i}^{-\frac{1}{2}} \mathbf{m}_i\left( \mathbf{x}_i - \mathrm{Dec}_\theta(\mathrm{Enc}_\phi( \mathbf{x}_i)) \right) \parallel_2^2
   \label{eq:NLL}
\end{equation}
where \(\mathbf{x}_i\) is the input image, \(\mathbf{\Sigma}_i \) is the diagonal noise covariance provided by the imaging pipeline, accounting for background and shot noise from bright sources, and $\mathbf{m}_i$ is the survey pipeline mask which removes masked pixels in the image. 

\subsubsection{Training Details}
\begin{figure}
    \centering
    \includegraphics[width=0.99\linewidth]{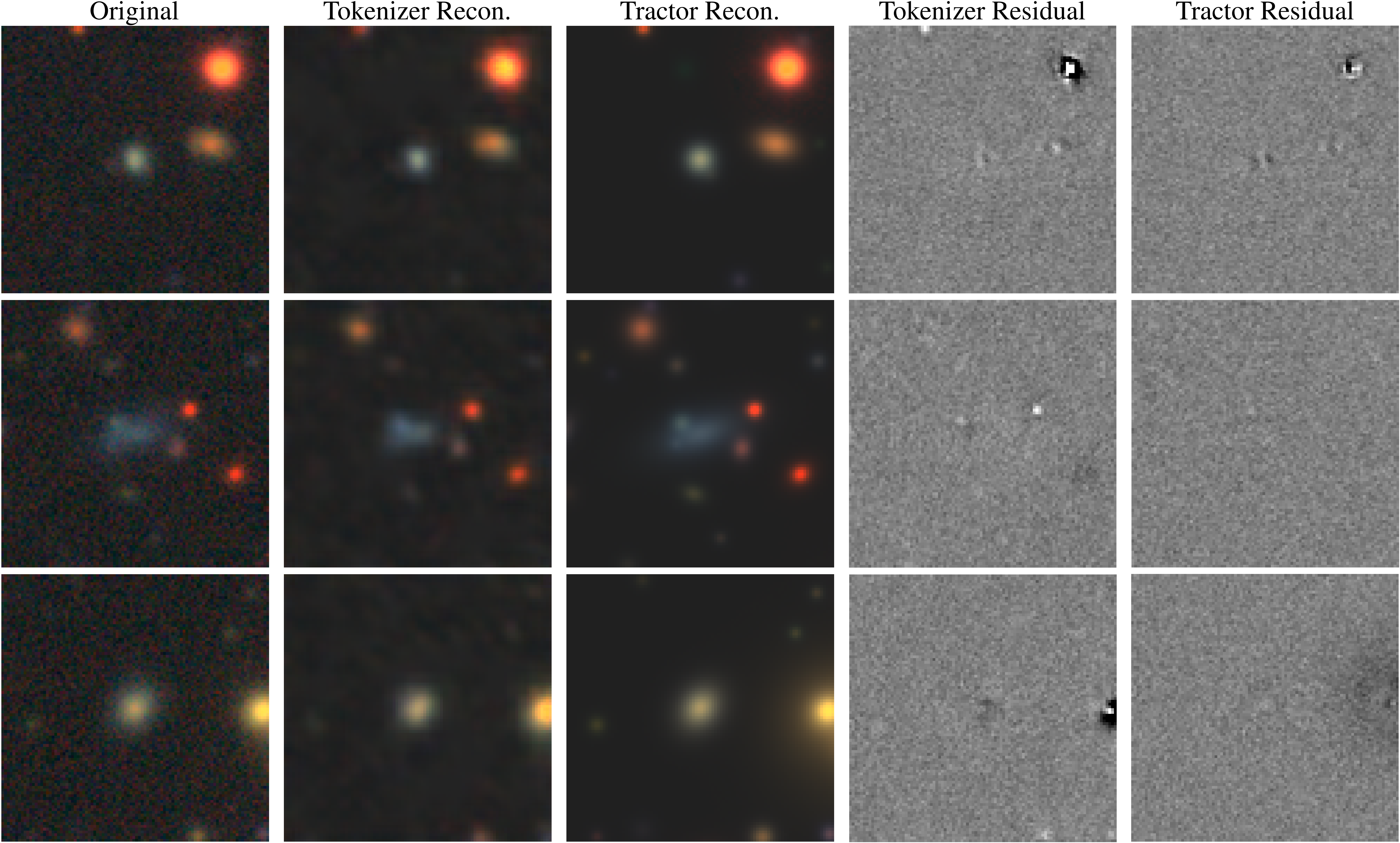}
    \caption{\textbf{Image Tokenizer Performance}: Reconstruction quality of the image tokenizer on three representative Legacy Survey images. The columns show, left to right, the original image, the reconstruction from the tokens, the reconstruction from the Legacy Survey tractor, the $r$-band residual for the tokenizer, and the $r$-band residual for the tractor.}
    \label{fig:image-autoencoding}
\end{figure}

We train with the image tokenizer using the Adam \citep{adamkingma} optimizer with a learning rate of $5\times10^{-4}$ on batches of 256 images, sampling LS:HSC at a 20:1 ratio to reflect the relative size of the two datasets. The learning rate is warmed up over 1k steps before being decayed for 400k steps using a cosine decay. Training converges in $\sim$5 days on 4 × NVIDIA H100 GPUs, yielding a final reconstruction score of $\mathcal{L}_\mathrm{NLL} = 0.00775$. We show some representative samples of the tokenizer's reconstruction quality in \autoref{fig:image-autoencoding}, and include reconstructions from the Legacy Survey pipeline tractor for comparison.

\subsection{Multi-Survey Spectrum Tokenizer} 
\label{spectrum_tokenizer}

\subsubsection{Preprocessing}
Each input spectrum provides (i) observed‐frame flux density per unit wavelength $\mathbf{f}(\lambda)$, (ii) inverse standard deviation $\mathbf{istd}(\lambda)$, and (iii) a per-pixel mask $\mathbf{m}(\lambda)$. Our tokenizer ingests heterogeneous measurements drawn from both DESI and SDSS. For each survey, we compute a robust median flux $\tilde{\mathbf{f}}$ (ignoring masked pixels), use a $\log_{10}$ range compression, and quantize it with a 1-D scalar tokenizer (codebook size = 1024). We then normalize $\mathbf{f}$ and $\mathbf{istd}$ by $\tilde{\mathbf{f}}$ and stack them into a 2-channel array.
The array is linearly interpolated onto a fixed latent grid of 8704 points covering $3500$–$10462.4$\AA\ at 0.8\AA\ spacing; this common grid is then shared between SDSS and DESI, and removes survey-specific wavelength/dispersion differences.

\subsubsection{Architecture \& Quantization}
\paragraph{Autoencoder} The stacked spectrum flux and inverse standard deviation $\mathbf{x}\in\mathbb{R}^{B\times2\times8704}$ is encoded by a 4-stage ConvNeXt-V2 backbone \citep{woo2023convnext}, consisting of an initial downsampling stack composed of a $4 \times 4$ convolution and LayerNorm, followed by three downsampling stacks of $2 \times 2$ convolutions and LayerNorms. Each of the four downsampling stacks is followed by multiple
ConvNeXt V2 processing blocks. This compresses the spectrum into a $273 \times 512$ latent space, which is then further downsampled to $10$ dimension before being fed to the quantizer; this dimensionality is chosen to conform to the $2^{10}$ codebook size used by the quantizer. Like with the image, these steps are inverted during the decoding part of the autoencoder.

\paragraph{Quantizer} We use a Look-up-Free Quantizer \citep[LFQ;][]{yu2023language} with an embedding dimension of ten (equating to a codebook size of 1024 codes) to convert the latent sequence into discrete codes. Contrary to the images, we find here here that LFQ quantization slightly outperforms FSQ for the spectrum.

\subsubsection{Losses}
For each spectrum we project the decoder output back to its native wavelength grid and apply three losses:
\begin{enumerate}
\item \textbf{Flux likelihood.} Gaussian NLL weighted by inverse variance $w(\lambda)$, identical to Eq. \eqref{eq:NLL}.
\item \textbf{Mask accuracy.} Binary cross-entropy between the predicted reliability map $\hat{\mathbf{m}}(\lambda)$ and the ground-truth mask $\mathbf{m}(\lambda)$.
\item \textbf{Commitment.} LFQ commitment loss with weight $\beta_{\mathrm{q}}=0.25$.
\end{enumerate}

\subsubsection{Training Details}
\begin{figure}
    \centering
    \includegraphics[width=0.99\linewidth]{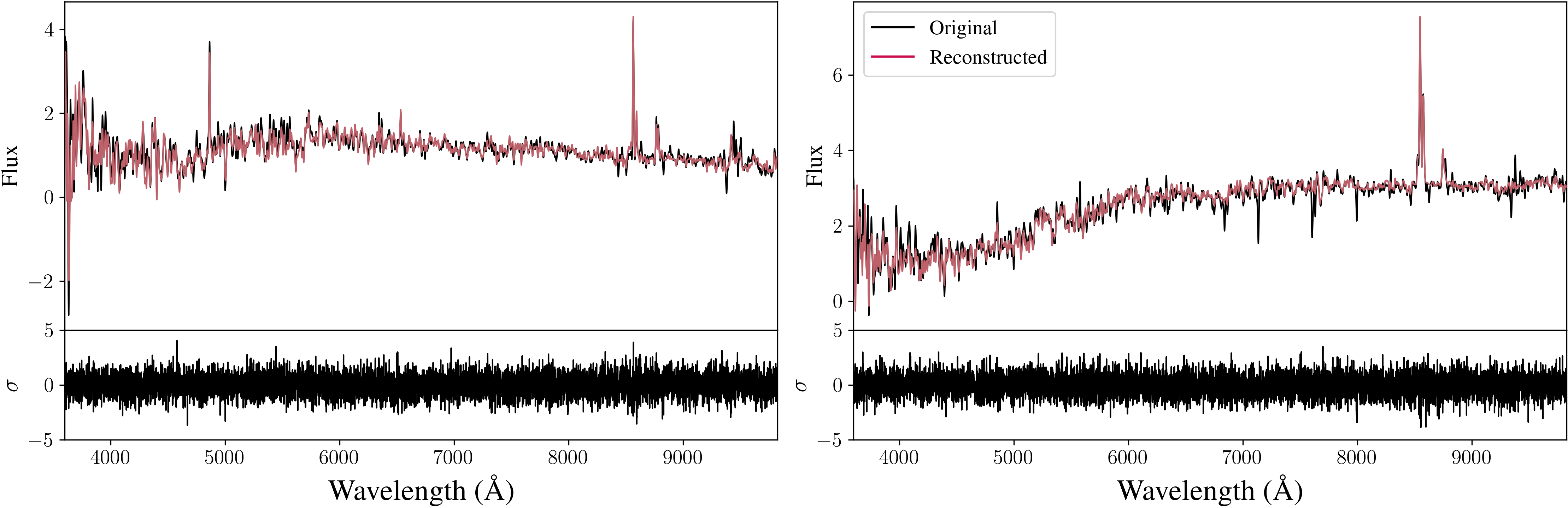}
    \caption{\textbf{Spectrum Tokenizer Performance}: Reconstruction quality of the spectrum tokenizer on two representative DESI EDR spectra. Top panels show the normalized flux as a function of wavelength, with original input (black) and reconstructed output (red) overlaid. Bottom panels show the residuals in units of the input uncertainty ($\sigma$).}
    \label{fig:spectrum-autoencoding}
\end{figure}

We train with using the AdamW optimizer with a constant $10^{-4}$ learning rate, a 0.1 weight decay penalty, and a global batch size of 128. Training for 215 k steps ($\sim$24 hours on 4 × NVIDIA H100) yields a token reconstruction $R^2=0.994$ and a mean mask AUC of 0.92. Reconstruction quality on two representative spectra from DESI are shown in \autoref{fig:spectrum-autoencoding}.

\subsection{Scalar Tokenizer}
\label{sec:scalar-tokenizer}

While the scalar values could be quantized directly, equal width binning directly in the data space would lead to an uneven probability mass assignment and potentially imbalance training. Therefore, we map every scalar value to a unit normal Gaussian before quantization. To that end, we first need to tabulate the empirical cumulative distribution function \(F_x\) on the training set\footnote{We estimate \(F_x\) with a fixed-size \emph{reservoir} (\(N\sim10^{6}\) samples) maintained online during data streaming. 
Reservoir sampling (Algorithm~R) produces an unbiased CDF while keeping memory \(\mathcal{O}(N)\), independent of the full catalog size.}. Once tabulated, we map a scalar value \(x_i\) to a standard normal variate via
\begin{equation}
    z_i \;=\; \Phi^{-1}\!\bigl(F_x(x_i)\bigr),
    \label{eq:scalar_gauss}
\end{equation}
where \(\Phi^{-1}\) is the inverse CDF of \(\mathcal{N}(0,1)\). Because \(z\sim\mathcal{N}(0,1)\), equal-width binning in \(z\)-space allocates the same probability mass to every bin, automatically adapting to long tails or sharp peaks in the original distribution. Each Gaussianised scalar \(z_i\) is quantised independently with an FSQ codebook of \(K=1024\) centroids. Centroids are fixed a priori: we place them at the \(K\) equally spaced quantiles of the standard normal, i.e.\ \(c_k = \Phi^{-1}\!\bigl((k-\tfrac12)/K\bigr)\).  
No parameters are learned and no loss is required. To recover an approximate scalar value \(\hat{x}_i\) from its token \(c_i\),
\[
    \hat{x}_i \;=\; F_x^{-1}\!\bigl(\Phi(c_i)\bigr).
\]
With \(K=1024\) bins the median absolute reconstruction error is below typical measurement uncertainties, ensuring that tokenisation fidelity is sufficient for downstream tasks while keeping the representation compact and parameter-free. Note that for some of the scalars with large dynamic ranges, we also apply a $\log_{10}$ or $\mathrm{arcsinh}$ transform before CDF mapping and tokenization. We apply the scalar tokenizer to the following scalars from each survey:

\begin{itemize}
    \item \textbf{Legacy Survey}: $\{g,r,i,z\}$ fluxes, WISE W1-W4 fluxes, $E(B-V)$ extinction, ellipticity components ($e_1,e_2$), circularized half-light radius $R_{\textrm{eff}}$.
    \item \textbf{HSC}: $\{g,r,i,z,y\}$ fluxes, shape tensor components.
    \item \textbf{SDSS \& DESI}: pipeline-reported redshift $(z)$.
    \item \textbf{Gaia}: 110 BP/RP coefficients, parallax, sky coordinates ($ra$, $dec$), $G$, BP, RP fluxes.
\end{itemize}

\subsection{Scalar Field Tokenizer}

In addition to images, we included an additional tokenizer specialized for scalar maps with values in $[0,1]$. This tokenizer is particularly adapted to handle segmentation maps, but could also be used to generate any property map scaled between 0 and 1, such as Star Formation Rate maps derived from Integral Field Spectroscopy. 

\subsubsection{Data \& Preprocessing}

The scalar field tokenizer was trained to autoencode a mixture of 5 categories of normalized single-channel images derived from Legacy Survey photometry: RGB cutouts converted to grayscale; individual red, green, and blue channels from the RGB cutouts; and an `object mask' indicating the silhouettes of sources detected in each cutout. The object mask is generated from the Tractor model photometry included in the Legacy Survey data release. The Tractor classifies each detected source as one of 5 morphological types and fits a corresponding elliptical surface brightness model to the light emitted by the source. After fitting, parameters can be extracted from the surface brightness profile to define a centered ellipse enclosing 50\% of the total emission from a given source. We generate an object mask for each cutout by painting such ellipses onto a null background for all sources detected in the cutout. The ellipses are filled with a constant value selected from $\{0.2, 0.4, 0.6, 0.8, 1.0\}$ which corresponds to the morphological type.

\subsubsection{Architecture \& Quantization}

\paragraph{Autoencoder} We base our architecture on VQ-VAE \citep{vq-vae}; the encoder is comprised of a stack of 3 convolutional downsampling layers followed by 2 residual blocks, and this arrangement is mirrored in the decoder (with upsampling transpose convolutions replacing the downsampling convolutions). The downsampling convolutions have kernel size 4, stride 2, padding 1, and 128 / 256 / 512 kernels, respectively. Each residual block consists of a sequence of 2 convolutional layers separated by a batch norm layer, where the convolutions have kernel size 3 / 1, stride 1, and padding 1 / 0, respectively. All layers are ReLU-activated.

\paragraph{Quantizer} We use a Finite Scale Quantizer \citep{mentzer2023finite} to quantize 4-dimensional codes with $n_{\text{dim}}=\{8, 5, 5, 5\}$ discrete levels available in the respective dimensions, yielding a codebook size of 1000.

\subsubsection{Training Details}

This tokenizer was optimized under a mean squared error objective using the AdamW optimizer with the weight decay parameter set to 0.01. The model was trained with batch size 256 for 114,000 steps at a base learning rate of $10^{-4}$. The base learning rate was modified by a linear warmup phase in the first 1,000 steps and a cosine decay over the final 72,000 steps. The model weights were updated as an exponential moving average of previous values with a decay parameter of 0.9999. Under these conditions, the loss converged to $\mathcal{L}_{\text{MSE}} \approx 0.0017$.

\section{Multimodal Masked Modeling}
AION-1 is inspired by many who came before us, and builds on recent early fusion multimodal models \citep{chameleonteam, perceiverio,imagebind}. In particular, it adopts the scalable multimodal masked modeling scheme proposed in 4M \citep{mizrahi20234m, 4m-21} to learn from heterogeneous data (e.g., spectra, images, scalars) by randomly masking tokenized inputs across all available modalities and reconstructing the masked content. 

Concretely, let \(\mathbf{X} = \{\mathbf{x}_1, \ldots, \mathbf{x}_M\}\) be token sequences for \(M\) modalities available for a training example. 
During training, two disjoint subsets of $\mathbf{X}$ are drawn: \(\mathbf{x}^{\mathrm{obs}}\) (observed) and \(\mathbf{x}^{\mathrm{tgt}}\) (target). Because these two subsets are sampled \emph{across} the entire token pool, the model learns both intra- and cross-modal relationships in each training instance.

The loss for the model is then given by:
\begin{equation}
        \mathcal{L}_{\mathrm{4M}}(\theta)
   \;=\;
    -\sum_{t=1}^{N} \log\,p_\theta\bigl(\mathbf{x_t}^{tgt} \,\big\vert\, \mathbf{x}_t^{\mathrm{obs}}\bigr),
\end{equation}
where \(p_\theta(\cdot \mid \mathbf{x}^{\mathrm{obs}})\) is the categorical distribution over the predicted vocabulary, and $N$ is the output token budget. 

\subsection{Architecture}

We adopt a Transformer-based \emph{encoder-decoder} framework suitable for the multi-modal masked prediction (see Figure~\ref{fig:main-figure}). Beyond a standard encoder-decoder architecture, we emphasize below the modality specific embedding scheme needed at the input of both the encoder and decoder to implement our objective.

Concretely, each modality \(i\in \{1,\dots,M\}\) has its own token embedding \(\mathrm{Embed}_i(\cdot)\), a learnable modality embedding \(\mathbf{m}_i\), and positional embedding \(\mathbf{p}_t\) for token position \(t\). Then, for an observed token from modality $i$, \(x_t^i\), the full embedding is given by 

\begin{equation}
      \mathbf{e}_t^{(\mathrm{enc})} 
      \;=\; 
      \mathrm{Embed}_i(x_t^i) 
      \;+\; 
      \mathbf{m}_i 
      \;+\;
      \mathbf{p}_t.
\end{equation}
In the decoder, we feed information on the \emph{target} we are querying tokens for, without providing their value:
\begin{equation}
       \mathbf{e}_t^{(\mathrm{dec})} 
       \;=\; 
       \mathbf{m}_i
       \;+\;
       \mathbf{p}_t,
\end{equation}
omitting any direct lexical embedding \(\mathrm{Embed}_i(x_t)\). 

In our implementation we use a different modality embedding \(\mathbf{m}_i\) for \textit{each modality} and \textit{each source} to identify the unique combination of data type and associated provenance metadata. In other words, two astronomical images from two different instruments will have two different modality embeddings even though they are both images. This is to provide the model with important provenance information which implicitly encodes aspects of data quality and resolution of the observations.

\subsection{Modality Masking Strategy}

A key consideration is to select which tokens become \emph{inputs} (observed) vs.\ \emph{outputs} (predicted) for each modality during training. We find that the dirichlet sampling from the original 4M implementation is ineficient when dealing with modalities that vary widely in length, and therefrore results in a high frequency of mostly empty batches. Therefore, we follow a simplified approach:

\paragraph{Input Token Budget} We select a global input token budget \(B\). To populate the budget, we first randomly pick one modality, and then uniformly randomly select a number of tokens for inclusion from that modality. We then fill the remaining budget $B$ by uniformly sampling tokens from the other modalities.

\paragraph{Output Token Budget.} For the remaining unselected tokens, we choose the number of tokens to predict for each modality by sampling from a Beta distribution skewed toward zero, which draws down the number of output tokens per sample, aligning with the eventual distribution of output tokens under a cosine schedule iterative sampling (e.g., in MaskGIT-style), ensuring that inference-time usage patterns are well covered during training. Similar to the input, one modality is chosen to draw an unconstrained number of tokens first, and the rest are filled by uniform random draws from the remaining modalities.

\section{AION-1 Family of Models}
\label{sec:aion-model}

\begin{figure*}[ht]
  \centering
  \begin{subfigure}[ht]{0.45\textwidth}
    \centering
    \includegraphics[width=\linewidth]{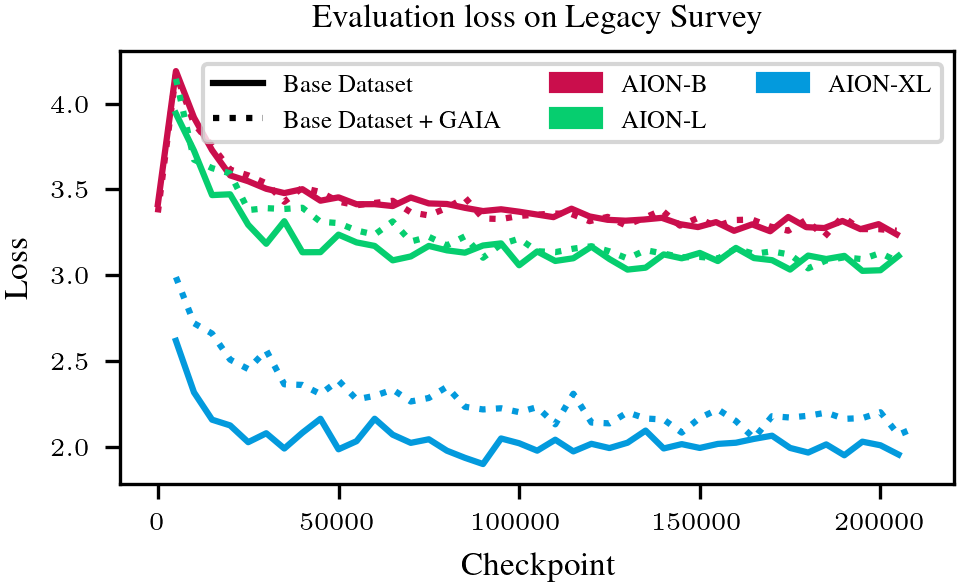}
    \label{fig:aion-eval-loss}
  \end{subfigure}\hfill
    \begin{subtable}[ht]{0.5\textwidth}
      \centering
      \small
      \label{tab:model-size}
      \vspace{0.8em}
      \begin{tabular}{@{}lccc@{}}
        \toprule
               & \textbf{AION-B} & \textbf{AION-L} & \textbf{AION-XL} \\ \midrule
        Enc. Blocks & 12   & 24    & 24   \\
        Dec. Blocks & 12   & 24 & 24 \\
        Dim    & 768  & 1024  & 2048 \\
        MLP Dim. & 3072 & 4096 & 16384 \\
        Heads  & 12   & 16    & 32   \\
        Params & 300M & 800M & 3B  \\ \bottomrule
      \end{tabular}
      \vspace{2.4em}
    \end{subtable}
  \label{fig:aion-loss-and-sizes}
  \caption{\textbf{AION-1 Scaling} \textit{Left}: Legacy Survey test losses for three model sizes, with and without the Gaia stellar set. Note the increase in loss with the inclusion of Gaia. \textit{Right}: AION-1 model variant breakdown of sizes, generally following the T5 model scaling convention \citep{raffel2020exploring}.}
\end{figure*}

We train three model versions - Base (300M), Large (800M), and XLarge (3B) - using the AdamW \citep{loshchilov2018decoupled} optimizer ($\beta_1=0.9, \beta_2=0.95$, weight decay $0.05$) for 205k steps with a global batch size of $8096$. We use a linear warmup and cosine decay schedule, with a peak learning rate of $2 \times 10^{-4}$. We adopt an input budget of 256 tokens, and output budget of 128 tokens for all our models during pretraining. All models are trained with bfloat16 mixed precision, and model distribution under PyTorch's Fully Sharded Data Parallel (FSDP) ZeRO-2 strategy. To achieve a batch size of 8192 in all cases, we train AION-1-B using 64 H100 GPUs for 1.5 days, AION-1-L using 100 H100 GPUs for 2.5 days, and AION-1-XL using 288 H100 GPUs for 3.5 days.

\section{Evaluation on Downstream Tasks}

This section evaluates AION-1's performance within an exemplary set of astronomical workflows. Ultimately, we aim to demonstrate that AION-1's streamlined foundation can significantly accelerate typical astrophysics tasks, facilitate data fusion, and provide superior results in low-data regimes, all while maintaining comparable or superior accuracy to typical supervised machine learning baselines. Although we present out-of-the-box generative capabilities, we propose using AION-1 primarily as a frozen backbone, as described in \cref{subsec:embeddings}.

\subsection{Out-of-the-Box Capabilities}

AION-1 is a generative model: once pretrained, it represents the joint distribution of all 39 tokenised modalities. At inference time we can therefore draw posterior samples of any modality in the training set by passing the appropriate query tokens and iteratively resampling them conditioned on the visible context. To generate these posteriors, we perform the following steps. 

First, we pass the query modality through its appropriate tokenizer, producing a set of \emph{input tokens} $\mathbf x^{\text{in}}\!=\!(x_1,\dots,x_{N_{\text{in}}})$. These are passed to the AION-1 encoder, while the decoder receives a sequence of \emph{query tokens}  $\mathbf x^{\text{qry}}\!=\!(x_{N_{\text{in}}+1},\dots,x_{N})$ whose values are to be inferred. At test-time we need samples from
\begin{equation}
   p_\theta\!\bigl(\mathbf x^{\text{qry}} \,\big|\, \mathbf x^{\text{in}}\bigr)
   \;=\;
   \prod_{j\in\mathcal Q}
   p_\theta\!\bigl(x_j \,\big|\, \mathbf x^{\text{in}}\bigr),
   \label{eq:posterior-cond}
\end{equation}
where $\mathcal Q$ indexes the query positions and $p_\theta$ is the categorical
distribution produced by the frozen decoder. To perform this sampling, we follow the ROAR generation scheme introduced in 4M \citep{mizrahi20234m}:
at each iteration $t$ we
\begin{enumerate}
  \item Draw a fresh \emph{random permutation}  
        $\pi_t : \mathcal Q_t \!\to\! \mathcal Q_t$ of the
        still-unknown query indices $\mathcal Q_t$;
  \item Reveal the first $\rho_t= \lfloor r^{\,t}\,|\mathcal Q_t|\rfloor$
        positions of this permutation,
        \[
          \mathcal S_t \;=\; \bigl\{ \pi_t(1),\dots,\pi_t(\rho_t) \bigr\};
        \]
  \item Sample those tokens once from the model,
        \[
          x_j^{(t)} \sim p_\theta\!\bigl(\,\cdot \,\big|\, \mathbf x^{\text{in}}\cup
          \mathbf x^{(t-1)}_{\mathcal Q_{t-1}\setminus\mathcal S_t}\bigr),
          \quad j\in\mathcal S_t;
        \]
  \item Promote them to inputs:
        $\mathbf x^{\text{in}}\!\leftarrow\!
        \mathbf x^{\text{in}}\cup \{x_j^{(t)}\}_{j\in\mathcal S_t}$ and
        update $\mathcal Q_{t+1} = \mathcal Q_t\setminus\mathcal S_t$.
\end{enumerate}
With a decay factor $r\!\in\![0,1)$ the number of unresolved tokens drops
exponentially, so the full sample is generated in $T=\mathcal O(\log|\mathcal Q|)$ decoder calls.  After sampling is complete, every query token is routed back through its modality-specific tokenizer to recover the original data representation. Repeating the entire ROAR loop $M$ times with a non-zero sampling
temperature $\tau>0$ yields $M$ i.i.d. draws $\{\hat{\mathbf x}^{(k)}\}_{k=1}^M$ from the conditional distribution in \autoref{eq:posterior-cond}.  

Importantly, we note that these draws are plausibility samples from the decoder’s categorical outputs under an iterative reveal schedule; they are not guaranteed to be well-calibrated joint posteriors for sequences of tokens longer than a single token. We discuss this limitation in further detail in \cref{subsec:limitations}.

\subsubsection{Redshift Estimation}

\begin{figure*}
    \centering
    \begin{subfigure}[b]{0.3\textwidth}
        \centering
        \includegraphics[width=\textwidth]{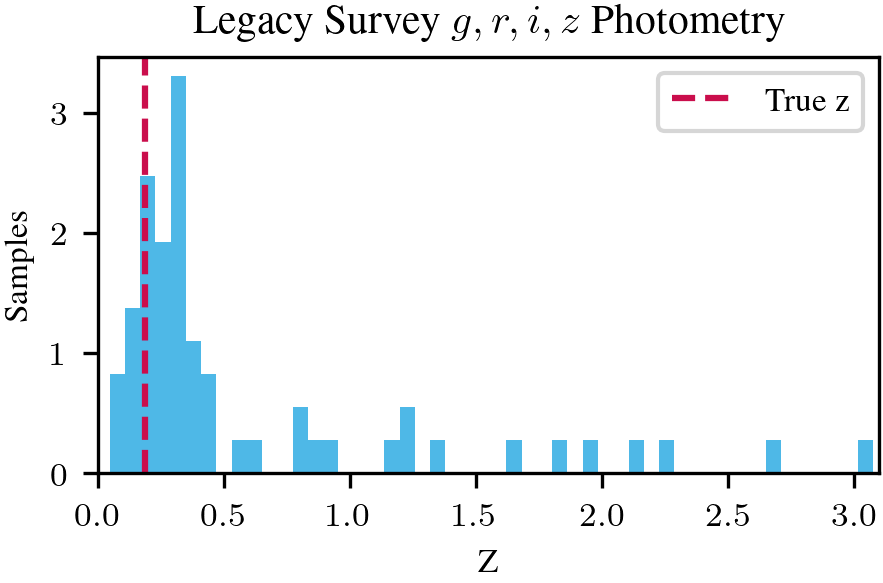}
    \end{subfigure}
    \hfill
    \begin{subfigure}[b]{0.3\textwidth}
        \centering
        \includegraphics[width=\textwidth]{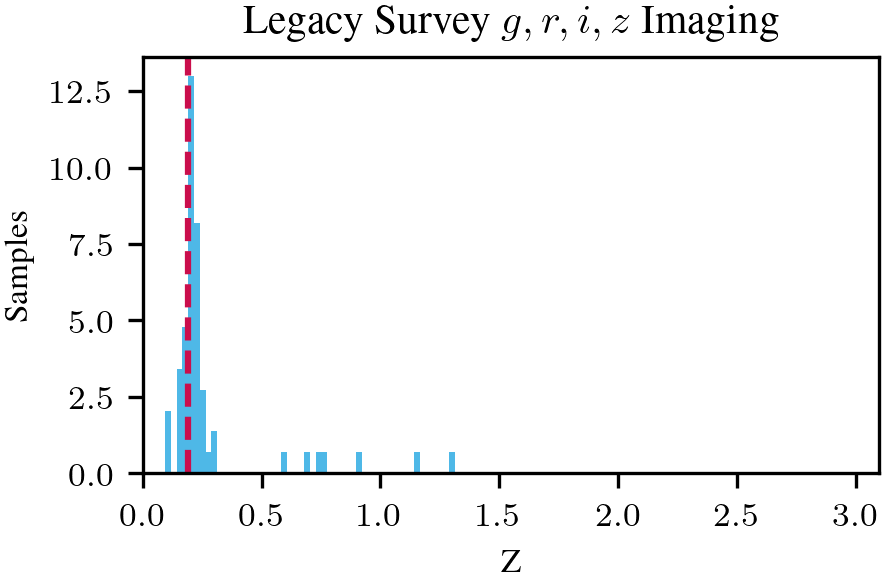}
    \end{subfigure}
    \hfill
    \begin{subfigure}[b]{0.3\textwidth}
        \centering
        \includegraphics[width=\textwidth]{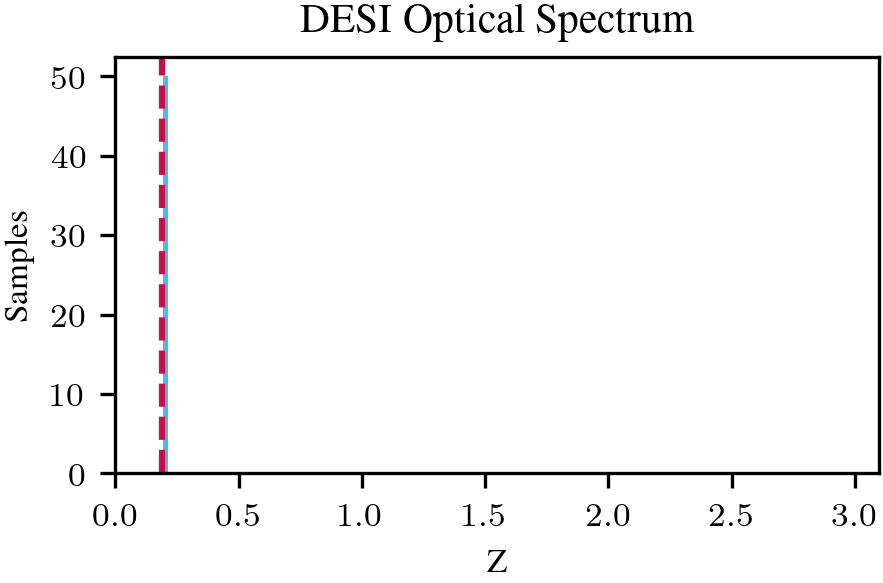}
    \end{subfigure}
    \caption{\textbf{Redshift Posterior Estimation}: Posterior samples for a single Legacy,Survey galaxy under three conditioning scenarios. \textit{Left}: broadband $g,r,i,z$ photometry alone yields a broad $p(z)$. \textit{Middle}: incorporating the corresponding $96\times96$ pixel multi-band cut-out (image$+$photometry) signigificantly tightens the credible interval. \textit{Right}: conditioning on the high-resolution DESI spectrum yields a perfect redshift estimate.}
    \label{fig:redshift}
\end{figure*}

Redshift $z$ is one of the scalar channels quantized during pretraining, so the decoder can naturally output a categorical distribution over the 1,024 quantized redshift bins. In this example, we tokenize an input modality, and using the scheme above, generate tokens (and corresponding redshifts) over $50$ ROAR draws for each modality. \autoref{fig:redshift} displays posterior samples for a representative galaxy under three increasingly informative contexts: (1) Legacy Survey $\{g,r,i,z\}$ photometry only, (2) Legacy Survey $\{g,r,i,z\}$ photometry and multi-band imaging, and (3) high-resolution DESI spectra. It is clear how the posterior contracts as richer information is provided: starting from broadband Legacy Survey photometry, adding spatial morphology from multi-band imaging, and finally a full optical spectrum.

\subsubsection{Spectral Super-Resolution}
\begin{figure}[ht]
    \centering
    \includegraphics[width=0.5\linewidth]{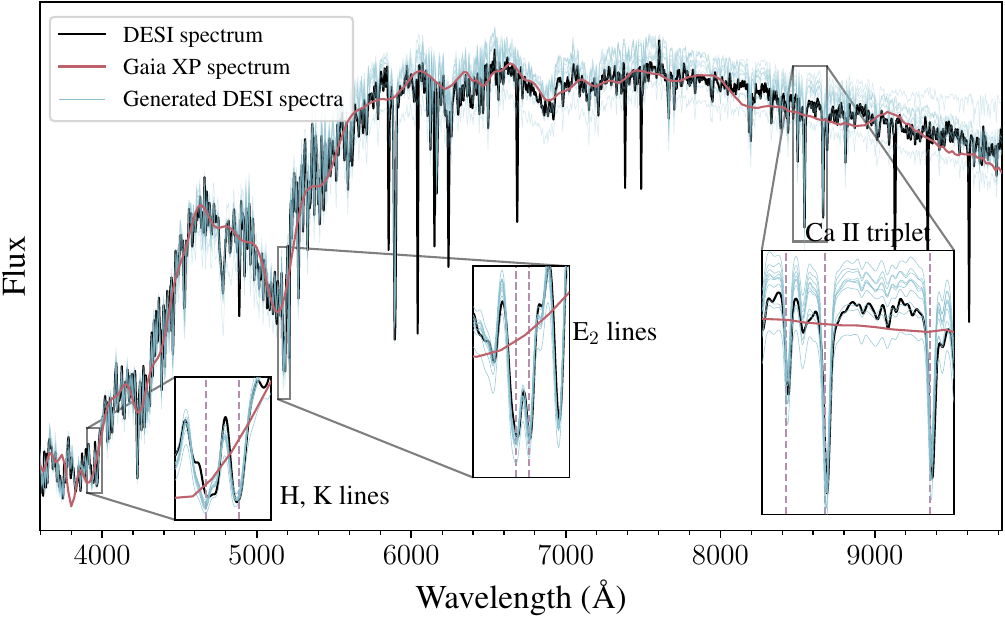}
    \caption{\textbf{Spectral Super-Resolution}: AION-1 can generate high-resolution posterior spectra samples (in this case, DESI; blue) conditioned on low-resolution spectra input (GAIA BP/RP coefficients; red), closely matching the ground-truth high-resolution measurements in line location, width, and amplitude. Several prominent lines are magnified in inset panels, with dashed lines marking their known locations.}
    \label{fig:conditional-generation}
\end{figure}

During pretraining, AION-1 learns to translate between multiple surveys, enabling conditional generation of one modality from another. Here, we demonstrate this capability by sampling query tokens corresponding to high-resolution DESI spectra while conditioning AION-1 on tokens from low-resolution GAIA BP/RP coefficients. Once sampled, the DESI tokens are passed through the spectrum decoder to produce realizations of the spectrum. \autoref{fig:conditional-generation} illustrates this generation on a representative star: the red curve is the native Gaia spectrum, the black curve is the true DESI measurement, and the blue posterior samples trace the super‐resolved features. AION-1 accurately recovers line centers, widths, and amplitudes within narrow posterior uncertainty bands, demonstrating its ability to impute high‐frequency spectral structure from coarse observations. 

Although the samples visually and quantitatively track key features, they should not be interpreted as calibrated joint posteriors over the full DESI spectrum; multi-token dependencies may be underrepresented by our current sampler as noted previously.

\subsection{AION-1 Embeddings}
\label{subsec:embeddings}


AION‑1’s primary practical benefit is its ability to produce powerful, physically meaningful, modality‑agnostic embeddings that work out of the box for a wide range of tasks, avoiding the engineering and data costs of end‑to‑end supervised pipelines. At the same time, because foundation models carry an implicit prior from pre‑training, we treat AION‑1 as a frozen feature extractor and perform lightweight, task‑specific calibration. In practice, we freeze the encoder, fit a small linear/MLP head on a representative calibration set that reflects the downstream selection function, and calibrate probabilities or continuous outputs. This workflow preserves AION‑1’s representational power while letting researchers inject the scientifically relevant prior and maintain control over inference.

\paragraph{Extracting embeddings.}  
To extract embeddings at inference time, we simply freeze the AION-1 encoder and discard the decoder.  Given the contextualized vector sequence
\(
\mathbf{Z}=\{\mathbf z_{1},\dots,\mathbf z_{T}\},\;
\mathbf z_{t}\in\mathbb{R}^{d},
\)
produced by the input modality fed through its appropriate tokenizer, we form an object-level vector $\mathbf e\in\mathbb{R}^{d}$ with one of two pooling schemes:

\begin{enumerate}
  \item \textbf{Mean pooling}:
  \begin{equation}
      \mathbf e \;=\; \frac{1}{T}\sum_{t=1}^{T}\mathbf z_{t}, 
  \end{equation}
  \label{eq:mean-pool}
  where the vectors are pooled deterministically by taking the average over all outbook vectors from the AION-1 encoder.

  \item \textbf{Attentive pooling}:
  \begin{equation}
      \mathbf A = \mathrm{softmax\ } \!\Bigl(\frac{\mathbf Q\mathbf K^{\!\top}}{\sqrt{d}}\Bigr),\qquad
      \mathbf e \;=\; \mathbf A\,\mathbf V,
  \end{equation}
  \label{eq:att-pool}
  where $\mathbf Q,\mathbf K,\mathbf V$ are learnable query/key/value projections applied to $\mathbf Z$, which are learned online during model adaptation to the downstream task.
\end{enumerate}

Both approaches yield a \emph{modality-agnostic} embedding: the same encoder handles 2-D images, 1-D spectra, or scalar tokens without architectural changes.  Moreover, AION-1 naturally supports multi-modality: one obtains a joint representation by concatenating the tokens from any subset of modalities and passing the union through the frozen encoder.  No extra fusion module is required—the cross-modal context was learned during pre-training. Ultimately, these embeddings encode astrophysically meaningful structure as we demonstrate in the rest of this section.

\subsubsection{Galaxy Parameter Estimation}

\begin{wraptable}{r}{0.55\textwidth}
\centering
\small
\begin{tabular}{l
                c  
                c  
                c  
                c  
                c  
               }
\toprule
 & $\mathbf{z}$ & $\mathbf{M_\star}$ & $\mathbf{t_{age}}$ & $\mathbf{\log Z_{\!Met}}$ & $\mathbf{sSFR}$ \\
\midrule
AION-1-B & & & & & \\
\hspace{1mm}\textit{Ph}            & 0.75 & 0.72 & 0.35 & 0.41 & 0.38 \\
\hspace{1mm}\textit{Ph+Im}         & 0.93 & 0.89 & 0.45 & 0.49 & 0.64 \\
\hspace{1mm}\textit{Ph+Im+Sp}      & 1.00 & 0.96 & 0.53 & 0.61 & 0.72 \\[2pt]
AION-1-L & & & & & \\
\hspace{1mm}\textit{Ph}            & 0.76 & 0.73 & 0.36 & 0.41 & 0.39 \\
\hspace{1mm}\textit{Ph+Im}         & 0.94 & 0.89 & 0.45 & 0.50 & 0.64 \\
\hspace{1mm}\textit{Ph+Im+Sp}      & 1.00 & 0.96 & 0.53 & 0.62 & 0.73 \\[2pt]
AION-1-XL & & & & & \\
\hspace{1mm}\textit{Ph}            & 0.79 & 0.76 & 0.31 & 0.38 & 0.48 \\
\hspace{1mm}\textit{Ph+Im}         & 0.94 & 0.89 & 0.45 & 0.49 & 0.64 \\
\hspace{1mm}\textit{Ph+Im+Sp}      & 0.99 & 0.95 & 0.53 & 0.62 & 0.73 \\
\midrule
\cite{Parker2024} & & & & & \\
\hspace{1mm}\textit{Im}$^{*}$      & 0.78 & 0.73 & 0.29 & 0.36 & 0.42 \\
\hspace{1mm}\textit{Sp}            & 0.99 & 0.90 & 0.52 & 0.60 & 0.70 \\[2pt]
\cite{oquab2023dinov2} & & & & & \\
\hspace{1mm}\textit{Im}$^{*}$      & 0.57 & 0.55 & 0.17 & 0.28 & 0.25 \\
\midrule
Supervised & & & & & \\
\hspace{1mm}\textit{Ph}$^{1}$      & 0.71 & 0.69 & 0.30 & 0.30 & 0.38 \\
\hspace{1mm}\textit{Im}$^{2}$      & 0.86 & 0.82 & 0.45 & 0.49 & 0.64 \\
\hspace{1mm}\textit{Sp}$^{3}$      & 1.00 & 0.85 & 0.43 & 0.62 & 0.68 \\
\bottomrule
\end{tabular}
\caption{\textbf{R$^2$ ($\uparrow$) for galaxy property estimation.}  
Inputs are photometry (\textit{Ph}), photometry + imaging (\textit{Ph+Im}), and photometry + imaging + spectra (\textit{Ph+Im+Sp}).  
$^{*}$AstroCLIP and DINOv2 use $\{g,r,z\}$ Legacy Survey images while AION-1 and supervised use $\{g,r,i,z\}$.  
Supervised models: $^{1}$XGBoost, $^{2}$ConvNeXt, $^{3}$Conv + Attention network.}
\label{tab:galaxy_properties_full}
\end{wraptable}

For quantitative evaluation we adopt the PRObabilistic Value‐Added Bright Galaxy Survey \citep[PROVABGS;][]{hahn2023desi}, which provides Bayesian spectral energy distribution (SED) fits for $\sim 140,000$ DESI Bright Galaxy Survey targets. We retain five global properties - redshift $z$, stellar mass $M_{\star}$, stellar–population age $t_{\mathrm{age}}$, gas‐phase metallicity $Z_{\mathrm{met}}$, and star‐formation rate ${\rm SFR}$ - and cross‑match the catalog against Legacy Survey DR10 $\{g,r,i,z\}$ imaging/photometry and DESI EDR spectra. Objects with $M_{\star}<0$ or non‐physical magnitudes are discarded, leaving $\sim 120,000$ galaxies. To stabilize the dynamic range we predict $\log Z_{\mathrm{met}}$ and $\log M_{\star}$ and convert ${\rm SFR}$ to the specific rate ${\rm sSFR}=\log({\rm SFR}/M_{\star})$.

Each input combination - photometry alone (\textit{Ph}), photometry $+$ imaging (\textit{Ph+Im}), and photometry $+$ imaging $+$ spectra (\textit{Ph+Im+Sp}) - is tokenised with the modality specific encoders described in \S\ref{sec:image_tokenizer}–\ref{sec:scalar-tokenizer}. The resulting token sequence (which is simply stacked in the case of multimodal inputs) is passed through the \emph{frozen} AION‑1 encoder and compressed with a single learned cross‑attention layer \eqref{eq:att-pool}; a lightweight, two-layer multilayer perceptron (MLP, hidden size = 256, GELU) then maps the $d$‑dimensional embedding to the target parameters. We quantify performance with the coefficient of determination,
$R^{2} \;=\; 1 \;-\; (\sum_{j}(y_{j}-\hat{y}_{j})^{2})/ (\sum_{j}(y_{j}-\bar{y})^{2})$, where $y_{j}$ are the ground-truth values, $\hat{y}_{j}$ the predicted values, and $\bar{y}$ the mean of the ground-truth sample.

We train the cross-attention layer and the MLP probe with mean‑squared error on $80\%$ of the cross-matched sample and report the coefficient of determination $R^{2}$ on the held‑out $20\%$ split. We also benchmark against three modality‑specific supervised networks trained end‑to‑end:
\begin{itemize}
\setlength\itemsep{2pt}
\item \textbf{XGBoost on photometry}—a boosted trees regressor using calibrated fluxes as features;
\item \textbf{ConvNeXt‐Tiny on images}—the \cite{MMU} vision baseline,  trained from scratch on $\{g,r,i,z\}$ Legacy Survey $96\times96$ cut‑outs;
\item \textbf{Conv$+$Attention on spectra}—a 1‑D CNN with gated attention pooling following \citet{melchior2023autoencoding} trained from scratch on the DESI optical spectra.
\end{itemize}
Additionally, we benchmark two strong self-supervised baselines.
First, AstroCLIP \citep{Parker2024}, a previous state-of-the-art multimodal foundation model for galaxies; we follow the authors’ recommended protocol, extracting frozen embeddings from the CLIP image encoder and training a lightweight MLP ontop of the embeddings. Note that AstroCLIP was trained on Legacy Survey $\{g,r,z\}$ cut-outs only, so in our setting it has access to one fewer band ($i$) than AION-1. Second, DIONv2 \citep{oquab2023dinov2} represents a widely used vision model; we feed RGB-converted $\{g,r,z\}$ images to the ViT-g/14 backbone and again attach the same MLP probe. Full results are presented in \autoref{tab:galaxy_properties_full}. We observe that across the board, AION-1 produces competitive results with minimal downstream adaptation required, and that its out-of-the-box multimodal fusion capabilities provide a powerful framework for downstream multimodal tasks.

\subsubsection{Galaxy Morphology Classification}
\label{sec:galaxy-morphology-classification}

We consider here the problem of classifying galaxy images into ten distinct morphology classes (e.g. spiral arms, merging galaxies) defined by Galaxy Zoo 10 \citep[GZ10;]{walmsley2022galaxy,Leung_2018gz10}. We construct the downstream sample by cross-matching the Galaxy Zoo 10 catalog with the Legacy Survey DR10 imaging footprint, yielding $\sim 8,000$ galaxies with $\{g, r, i, z\}$ cutouts. For AION-1, we tokenize each cutout with the multi-survey image tokenizer, mean-pool the resulting embeddings\footnote{Although we experiment with attentive pooling in this setting, unlike with property estimation, we find that attentive pooling does not provide any meaningful gain in accuracy.} as in \autoref{eq:mean-pool}, and pass the 768-d mean vector to a two-layer MLP head (hidden size = 256, GELU, dropout = 0.1). The head is trained on $80\%$ of the sample with class-stratified splits and evaluated on the remaining $20\%$.

\begin{wraptable}{l}{0.45\textwidth}
  \centering
  \small
  \begin{tabular}{lc}
    \toprule
    \textbf{Model} & \textbf{Accuracy (\%)} \\
    \midrule
    AION-B        & 84.0 \\
    AION-L        & 87.2 \\
    AION-XL       & 86.5 \\
    \midrule
    \cite{oquab2023dinov2}        & 71.4 \\
    \midrule
    EfficientNet  & 80.0 \\
    \cite{walmsley2022galaxy}        & 89.6 \\
    \bottomrule
  \end{tabular}
  \caption{\textbf{Galaxy Morphology Classification Accuracy ($\uparrow$)} on Galaxy Zoo 10. AION-1, DINOv2 \citep{oquab2023dinov2}, and ZooBot \citep{walmsley2022galaxy} use an MLP head on frozen embeddings; EfficientNet-B3 \citep{tan2019efficientnet} is trained end-to-end from scratch.}
  \label{tab:wrap_acc}
\end{wraptable}

We replicate this protocol with the DINOv2 baseline, replacing the tokenizer with the ViT-g/14 backbone and applying the RGB normalization recommended by \citet{oquab2023dinov2}. EfficientNet-B3 is trained end-to-end from random initialization using the same splits and standard data augmentations. Finally, we adapt ZooBot \citep{walmsley2022galaxy} by fine-tuning the penultimate layer on our $8,000$ samples; although ZooBot was never exposed to GZ10 labels, it benefits from pre-training on $\sim 300,000$ images covering the broader, harder GZ-5 decision tree, and thus acts as an approximate upper bound on achievable accuracy.

As Table \ref{tab:wrap_acc} shows, AION-1-L tops all baselines except ZooBot, exceeding EfficientNet by $+7.2$ pp and DINOv2 by $+15.8$ pp, while using only a lightweight MLP head. Moreover, it reaches close to the ZooBot accuracy, only under-performing by $-2.4$ pp, despite seeing two orders of magnitude fewer labeled galaxy images during its inference process.  

\subsubsection{Galaxy Image Segmentation}
\label{sec:galaxy-image-segmentation}

\begin{figure}[ht]
  \centering
  \begin{subfigure}[t]{0.56\textwidth}
    \centering
    \includegraphics[width=\linewidth]{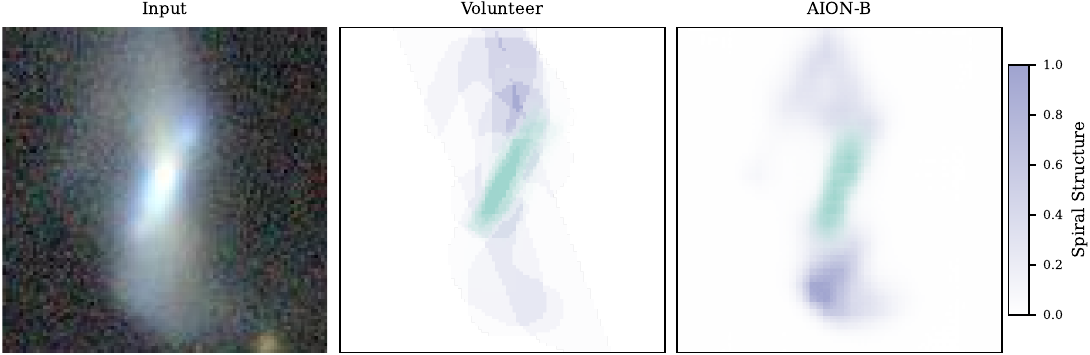}\\[2mm]
    \includegraphics[width=\linewidth]{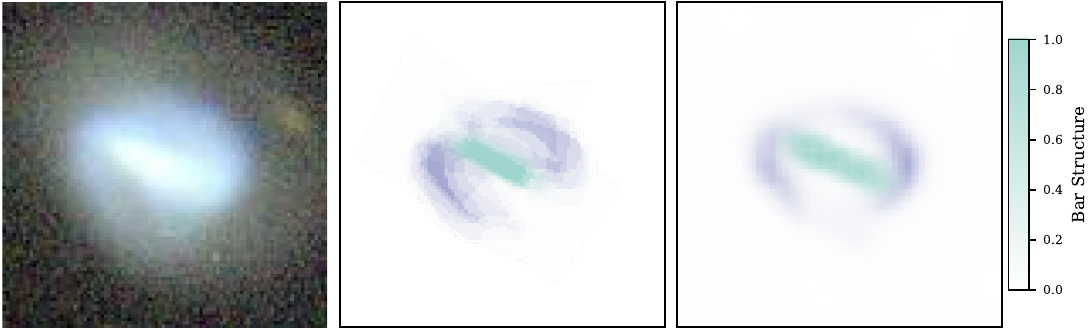}
    \label{fig:segmentation_examples}
  \end{subfigure}%
  \hfill
  \begin{subfigure}[t]{0.44\textwidth}
    \centering
    \small
    \begin{tabular}{lcc}
      \toprule
                          & \textbf{Spiral Arms} & \textbf{Bar} \\
      \midrule
      AION-1-B   & 0.60 & 0.31 \\
      AION-1-L  & 0.61 & 0.31 \\
      AION-1-XL  & 0.61 & 0.32 \\
      \midrule
      U-Net      & 0.55 & 0.28 \\
      \bottomrule
    \end{tabular}
    \label{tab:segmentation_iou}
  \end{subfigure}
  \caption{\textbf{Galaxy structure segmentation.}  
           \textit{Left}: Examples of galaxy image segmentation produced from the AION-1 embeddings with a lightweight convolutional neural network compared with the ground-truth true volunteer labels.
           \textit{Right}: Mean IoU ($\uparrow$), given by \autoref{eq:iou}, on the held-out test set.}
  \label{fig:segmentation_qual_quant}
\end{figure}

Going beyond broad morphology classification, we consider image segmentation based on human annotation of prominent galaxy structures obtained through the Galaxy Zoo 3D citizen science campaign \citep{10.1093/mnras/stab2282}. As above, we cross-match the sample with the Legacy Survey images, producing roughly 2,800 galaxy image-annotation pairs. We feed the $\{g,r,i,z\}$ cut-outs through the AION-1 model to produce embeddings using the mean-pooling\footnote{As with galaxy morphology classification, we find no benefit from attentive pooling relative to mean pooling.} aggregation scheme as in \autoref{eq:mean-pool}. Then, we train a lightweight convolutional upsampler to predict the image-level segmentation maps from the AION-1 embeddings. Our upsampler design is largely inspired by the mask decoder from \citet{sam}\footnote{\url{https://github.com/facebookresearch/segment-anything/}}, but with a key modification: we do not include hypernetworks instantiated from additional register tokens. Instead, we use a single convolutional layer to project the upsampled output to the desired number of segmentation maps, simplifying the architecture while maintaining efficiency. Training is performed on $80\%$ of the dataset, while $20\%$ is held-out for validation. Additionally, we train a supervised, end-to-end U-Net baseline directly on the image-segmentation pairs, following the architecture from \citep{walmsley2023deeplearningsegmentationspiral}. We measure performance on the held-out test set with the standard intersection-over-union  
\begin{equation}
    \mathrm{IoU}
  \;=\;
  \frac{\lvert M_{\text{pred}}\cap M_{\text{true}}\rvert}
       {\lvert M_{\text{pred}}\cup M_{\text{true}}\rvert},
\end{equation}
\label{eq:iou}
averaged over images containing the structure. We present both the IoU on the held-out test set, as well as sample AION-1-B segmentations, in \autoref{fig:segmentation_qual_quant}. AION-1’s frozen encoder plus a small decoder outperforms the supervised U-Net by $+0.06$ IoU on spiral arms and $+0.04$ on bars.

\subsubsection{Stellar Parameter Estimation}

To evaluate AION-1 on stellar parameters we assemble a heterogeneous testbed that combines \textit{Gaia} DR3 photometry and low-resolution XP spectra with high-resolution DESI optical spectra. Starting from the $\sim45$ M stars used during pre-training, we cross-match them against the DESI EDR and retain only those sources that (i) fall inside the pre-training \textit{validation} \textsc{healpix} tiles—thereby guaranteeing that AION-1 has \emph{never} seen them before—and (ii) possess reliable stellar labels from the catalogue of \citet[][hereafter Z24]{Zhang2024}.
Z24 employs a data-driven regularised model to infer $T_\mathrm{eff}$, $\log g$, [Fe/H], micro-turbulent velocity $v_\mathrm{mic}$, and 18 elemental abundances directly from DESI spectra; here we focus on the four global parameters shared with the literature baselines.
The resulting sample contains $\sim240,000$ stars, of which $80\%$ are used to train downstream heads and $20\%$ are reserved for evaluation.

Each input configuration—photometry alone (\textit{Ph}); \textit{Ph} augmented with parallax and sky position (\textit{Ph + Plx + RA/Dec}); \textit{Gaia} XP spectra plus the previous channels (\textit{XP + Ph + Plx + RA/Dec}); and DESI spectra with parallax (\textit{Sp + Plx})—is processed by the corresponding modality-specific encoders (\autoref{spectrum_tokenizer}–\autoref{sec:scalar-tokenizer}) to yield a single, mixed-modality token stream. Following the protocol adopted for galaxy properties, we pass the tokens through the frozen AION-1 encoder, apply a single learned cross-attention pooling layer \autoref{eq:att-pool} that compresses the variable-length sequence to a fixed $d$-dimensional vector; and map this vector to the target stellar labels with a linear probe. 

In addition to the AION-1 results, we also train two baselines end-to-end for comparison:
\begin{itemize}
    \item \textbf{ConvNeXt regressor on raw spectra.} A stack of ConvNeXt–Tiny blocks and down-sampling layers identical to the AION-1 spectral encoder is trained end-to-end on DESI spectra and inverse-variance arrays, followed by gated attention pooling and a linear head which predicts the stellar properties.
    \item \textbf{XGBoost on token representations.} We mean-pool either (1) the 1-D Gaia photometry tokens or (2) the 1-D Gaia photometry tokens and XP coefficients and the DESI spectra produced by the frozen AION-1 encoder and fit a gradient-boosted decision-tree regressor to predict stellar properties, following the same general procedures as in \cite{Leung2024}.
\end{itemize}

\begin{wraptable}{r}{0.5\textwidth}
  \centering
  \small
  \begin{tabular}{lcccc}
    \toprule
                       & $\mathbf{T_{\!eff}}$ & $\mathbf{\log g}$ & $\mathbf{[Fe/H]}$ & $\mathbf{v_{\mathrm{mic}}}$ \\
    \midrule
    \multicolumn{5}{@{}l}{AION-1-B}\\[-3pt]
    \hspace{1mm}\textit{Ph}                       & 0.94 & 0.95 & 0.58 & 0.86 \\
    \hspace{1mm}\textit{Ph+Plx+RA/Dec}            & 0.94 & 0.95 & 0.70 & 0.87 \\
    \hspace{1mm}\textit{XP+Ph+Plx+RA/Dec}         & 0.96 & 0.98 & 0.91 & 0.89 \\
    \hspace{1mm}\textit{Sp+Plx}                   & 0.99 & 0.98 & 0.94 & 0.89 \\[2pt]

    \multicolumn{5}{@{}l}{AION-1-L}\\[-3pt]
    \hspace{1mm}\textit{Ph}                       & 0.95 & 0.96 & 0.58 & 0.87 \\
    \hspace{1mm}\textit{Ph+Plx+RA/Dec}            & 0.95 & 0.96 & 0.71 & 0.88 \\
    \hspace{1mm}\textit{XP+Ph+Plx+RA/Dec}         & 0.97 & 0.98 & 0.92 & 0.89 \\
    \hspace{1mm}\textit{Sp+Plx}                   & 0.99 & 0.98 & 0.94 & 0.89 \\[2pt]

    \multicolumn{5}{@{}l}{AION-1-XL}\\[-3pt]
    \hspace{1mm}\textit{Ph}                       & 0.92 & 0.94 & 0.56 & 0.85 \\
    \hspace{1mm}\textit{Ph+Plx+RA/Dec}            & 0.93 & 0.95 & 0.68 & 0.87 \\
    \hspace{1mm}\textit{XP+Ph+Plx+RA/Dec}         & 0.97 & 0.97 & 0.91 & 0.88 \\
    \hspace{1mm}\textit{Sp+Plx}                   & 0.98 & 0.98 & 0.92 & 0.89 \\
    \midrule
    \multicolumn{5}{@{}l}{XGB Baseline}\\[-3pt]
    \hspace{1mm}\textit{Ph}$^{1}$                 & 0.94 & 0.95 & 0.59 & 0.87 \\
    \hspace{1mm}\textit{XP+Ph+Sp}$^{1}$           & 0.99 & 0.98 & 0.89 & 0.89 \\[2pt]

    \multicolumn{5}{@{}l}{ConvNeXT Baseline$^{2}$}\\[-3pt]
    \hspace{1mm}\textit{Sp}                       & 0.99 & 0.98 & 0.95 & 0.89 \\
    \bottomrule
  \end{tabular}
  \caption{\textbf{$R^2$ ($\uparrow$) for stellar-label prediction.}  
           Inputs: Gaia photometry (\textit{Ph}), low-resolution Gaia XP spectra (\textit{XP}),  
           parallax (\textit{Plx}), celestial coordinates (\textit{RA/Dec}), and high-resolution DESI spectra (\textit{Sp}).  
           $^{1}$Gradient-boosted trees (XGBoost).  
           $^{2}$Baseline convolution–attention network trained on spectra only.}
  \label{tab:stellar_properties_full}
\end{wraptable}

For all models, we minimize MSE on the training split and report the coefficient of determination $R^{2}$ on the held-out set.
Results for all three AION-1 sizes, together with supervised and self-supervised baselines, are summarized in \autoref{tab:stellar_properties_full}. AION-1’s multimodal fusion yields state-of-the-art performance with minimal adaptation: adding geometric information (\textit{Plx+RA/Dec}) noticeably improves metallicity, while incorporating XP spectra delivers a further $\sim$15–20\% absolute gain in [Fe/H] $R^{2}$. When high-resolution DESI spectra are available, AION-1 matches the supervised ConvNeXt baseline despite using \emph{frozen} weights, underscoring the quality of its spectral representations.
Overall, the model’s ability to ingest heterogeneous inputs and deliver consistently strong predictions highlights its potential as a cross-survey foundation for stellar astrophysics.

In addition to the supervised baselines presented above, we also compare performance on the stellar parameter regression task with a current state-of-the-art baseline from \cite{Leung2024}, who developed a Transformer-based foundation model for stellar data. More specifically, the task is to predict APOGEE-derived stellar parameters - namely $T_{\rm eff}$, $\log g$, and [Fe/H] - from Gaia XP spectral coefficients. We use the same data as \cite{Leung2024}, and cross-match APOGEE-derived stellar parameters with the MMU Gaia data, producing a set of roughly $\sim 10,000$ APOGEE parameter-Gaia XP spectral pairs. We feed as input to both AION-1 and the \cite{Leung2024} model only the first 32 BP coefficients and first 32 RP coefficients due to the fact that the \cite{Leung2024} model only has a context length of 64; we artificially handicap AION - which does not have this restriction - in order to perform a fair comparison. We note here that \cite{Leung2024} has been explicitly given APOGEE-derived stellar parameters and Gaia XP coefficients during its pretraining stage, and so this task is one that it has effectively been trained for. On the other hand, the pretraining dataset for AION-B does not contain APOGEE data, nor any other stellar parameters; we simply train a simple linear projection layer with cross-attention pooling on 5000 paired examples, and leave the weights of the AION model itself frozen. Despite this, we outperform the \cite{Leung2024} model across all parameters, as shown in \autoref{tab:stellar_properties_comparison}.

\begin{table}[h]
  \centering
  \small
  \begin{tabular}{lccc}
    \toprule
    \textbf{Model} & $T_{\rm eff}$ (K) & $\log g$ (dex) & [Fe/H] (dex) \\
    \midrule
    AION-B            & 94.6 & 0.206 & 0.115 \\
    \cite{Leung2024}  & 99.1 & 0.229 & 0.143 \\
    \bottomrule
  \end{tabular}
  \caption{\textbf{Performance on APOGEE stellar property predictions from Gaia XP coefficients} \textbf{($\downarrow$)}, as measured by standard deviation of residuals. K is temperature units in Kelvin, and dex represents scatter on a logarithmic scale.}
  \label{tab:stellar_properties_comparison}
\end{table}

\subsubsection{Performance in Low-Data Regime}

\begin{figure*}[h]
    \centering
    \begin{subfigure}[b]{0.3\textwidth}
        \centering
        \includegraphics[width=\textwidth]{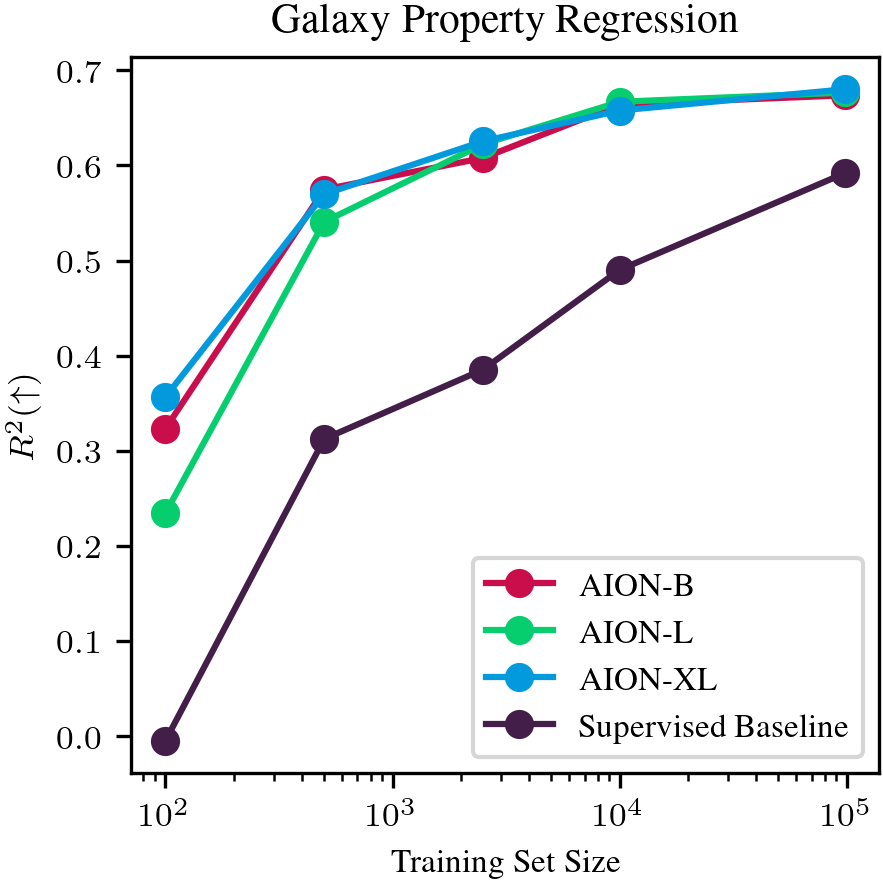}
        \caption{Galaxy Properties}
        \label{fig:sub1}
    \end{subfigure}
    \hfill
    \begin{subfigure}[b]{0.3\textwidth}
        \centering
        \includegraphics[width=\textwidth]{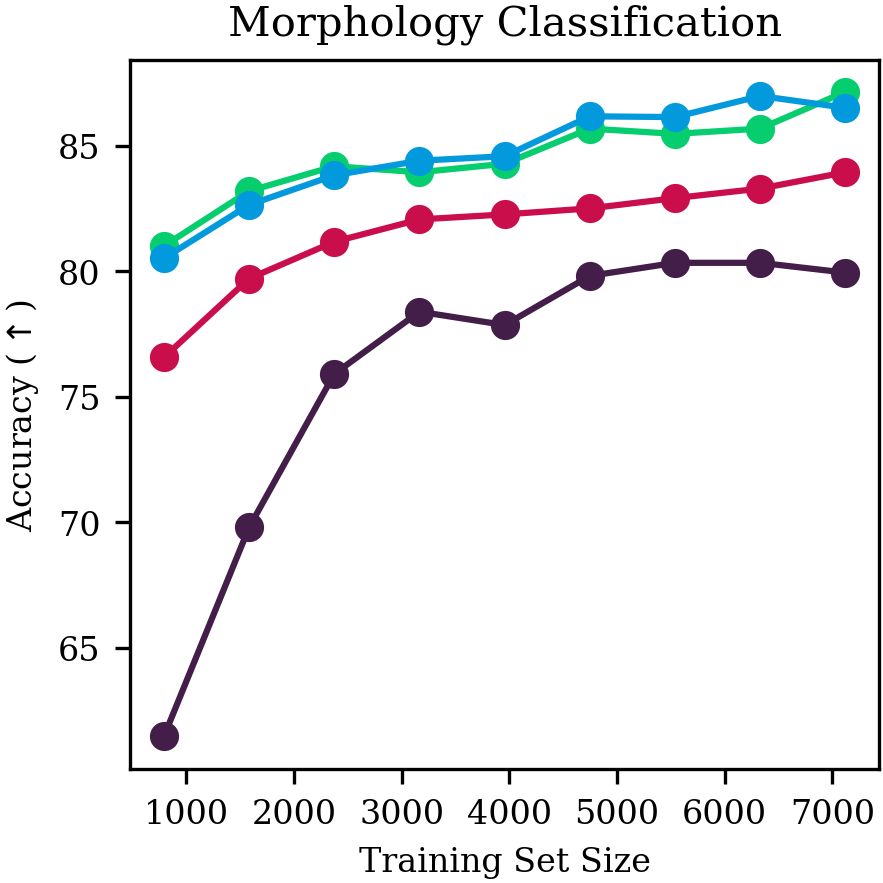}
        \caption{Morphology Classification}
        \label{fig:sub2}
    \end{subfigure}
    \hfill
    \begin{subfigure}[b]{0.3\textwidth}
        \centering
        \includegraphics[width=\textwidth]{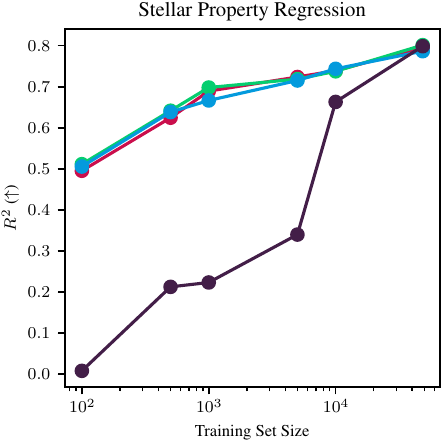}
        \caption{Stellar Properties}
        \label{fig:sub3}
    \end{subfigure}
    \caption{\textbf{Model performance vs downstream task training set size}. We regress \textbf{(a)} galaxy physical properties from images, \textbf{(b)} classify galaxy morphology, and \textbf{(c)} regress stellar properties from spectra on the same held-out test sample equating to $20\%$ of the available data. However, we artificially reduce the training set size, and train a lightweight head on top of the frozen AION-1 encoder and a supervised model on the raw input data for each training set size. For galaxy and stellar properties, we report the $R^2$ averaged over all the properties, while for morphology classification we report the average accuracy.}
    \label{fig:scaling_figures}
\end{figure*}

In astronomy, many key classes (e.g.\ rare transients, metal-poor stars, high-redshift galaxies) come with only a handful of reliable labels. As a result, approaches that continue to perform well in the low-data regime are particularly valuable. We explore AION-1's performance in this setting across three of the tasks described above: galaxy parameter estimation from images, stellar parameter estimation from spectra, and galaxy morphology classification from images. To that end, for these experiments, we keep the test set fixed at 20\% of the total available data volume, but artificially reduce the size of the available training data. For each subset of training data, we retrain the lightweight head on top of the frozen AION-1 encoder as well as the supervised baselines from scratch. For galaxy and stellar tasks we average $R^{2}$ across the target parameters; for morphology we report overall accuracy. Figure \autoref{fig:scaling_figures} shows two encouraging trends:
\begin{enumerate}
    \item \textbf{Low-Data Performance}: With only $10^{2}\!-\!10^{3}$ labels, AION-1 already reaches $R^{2}\!\sim\!0.5$ for physical properties and $\geq 80\%$ for galaxy morphology, while end-to-end supervised models remain near-zero $R^{2}$ or below $70\%$ accuracy, highlighting AION-1's performance even in extremely low-data regimes.
    \item \textbf{Faster Data Saturation} While AION-1 plateaus by $\sim 10^{3}\!-\!10^{4}$ training examples; the baselines need an order of magnitude more data to approach similar performance.
\end{enumerate}
Together, these results demonstrate that a strong, multimodal foundation model can transfer rich physical knowledge into downstream tasks long before a sizable, task-specific training set exists. In practical terms, observatories and survey teams can deploy a frozen AION-1 backbone and achieve competitive results with only a few hundred carefully vetted labels—orders of magnitude fewer than conventional supervised pipelines require. For future tasks, this may reduce the need for expensive spectroscopic follow-up or large volunteer-led annotation campaigns and make it feasible to apply machine learning to data-sparse niches. 

\subsection{Rare Object Detection}
\label{sec:retrieval}

Modern astronomical surveys catalog hundreds of millions of sources, however, the discoveries that most advance cosmology often lie in the distribution’s extreme tail—for example, strong gravitational lenses that constrain dark-matter substructure and cosmic expansion. Because such phenomena appear so infrequently, assembling large, well-annotated training sets is inherently difficult. The small catalogs that do exist are usually built with hand-tuned selection cuts, baking those choices—and their attendant biases—into any supervised model trained on them. As a result, effective search for rare objects requires methods that minimize reliance on task-specific labels and remain robust to selection bias. AION-1 addresses this challenge by \emph{semantic retrieval} in latent space, in which AION-1 is used to embed both \emph{query} galaxies and all \emph{candidate} images into a shared space.  Cosine similarity then ranks the corpus, enabling zero‑shot discovery without specialist tuning.

\paragraph{Set-Up} Specifically, given a query galaxy-which, for example, could be a known strong gravitational lens-AION-1 is used to produce an embedding of the galaxy's image, $\mathbf{z_q} \!\in\! \mathbb{R}^{d}$; in this case, we use mean-pooling over AION-1's output vectors, as in \autoref{eq:mean-pool}. For all other galaxies in a search corpus, which we will refer to as candidates, a series of candidate embeddings, $\mathbf{z_c}$, are also produced. For each candidate, we compute the cosine similarity between its embedding and the query embedding, given by
\begin{equation}
\label{eq:cosine-similarity}
S_c(\mathbf{x_q}, \mathbf{x_c}) \;=\;
\frac{\mathbf{x_q}^{\mathsf T}\mathbf{x_c}}
     {\|\mathbf{x_q}\|_2\;\|\mathbf{x_c}\|_2}.
\end{equation}
All candidates can then be sorted in descending order of $S_c$, and the top $N$ galaxies can be returned as the most suitable candidates for that type of query. 

In order to quantify the performance of AION-1 on such a task, we use the normalized Discounted Cumulative Gain (nDCG). Specifically, let $r_i$ be the relevance label of the candidate object ranked at position $i$.  The discounted cumulative gain (DCG) at rank $k$ is
\begin{equation}
\mathrm{DCG@}k \;=\;
\sum_{i=1}^{k} \frac{2^{r_i}-1}{\log_2(i+1)}.\label{eq:dcg}
\end{equation}
The ideal DCG ($\mathrm{IDCG@}k$) is computed with the candidates ordered by descending relevance; finally
\[
\mathrm{nDCG@}k \;=\; \frac{\mathrm{DCG@}k}{\mathrm{IDCG@}k}.
\]
We adopt $k=10$ in all experiments.

\paragraph{Galaxy‑Zoo Retrieval Experiments} We construct a benchmark from the Galaxy Zoo–DECaLS catalog \citep[GZ-DECaLS;][]{walmsley2022galaxy}, which provides citizen-science morphology votes for Legacy Survey galaxies \citep{dey2019overview}. After removing objects with fewer than three volunteer votes and cross-matching the GZDECaLS with the Legacy Survey SGC, the sample contains $\sim$171\,000 galaxies. We focus on two visually distinctive classes—\emph{mergers} and \emph{spirals}, and focus on high-confidence exemplars to form our query galaxies, which we define as galaxies for which the fraction of volunteers who agreed on the class $f$ exceeds $90\%$. This yields roughly 700 merger queries and 25,000 spiral queries. For each query (merger or spiral), we then return the top $k=10$ candidate galaxies according to \autoref{eq:cosine-similarity}, and compute the $\mathrm{nDCG@}10$ score, where the relevance score of the retrieved galaxies $r_i$ is given by the vote fraction for the query class; e.g., a galaxy for which $70\%$ of the volunteers labeled the galaxy a spiral receives a relevance score of $r\!=\!0.7$ for a spiral query. This soft-label strategy rewards the retrieval of unambiguous examples while still giving partial credit to visually ambiguous cases. Ultimately, the nDCG scores are averaged over all query vectors and reported for that class.

\begin{table}
  \centering
  \small
  \begin{tabular}{l l l r c}
    \toprule
    Type   & Dataset          & Image Type    & Total Number & Frequency in Dataset \\ 
    \midrule
    Spiral & GZ-DECaLS      & Legacy Survey      & 24\,622 & 26\% \\
    Merger & GZ-DECaLS     & Legacy Survey       & 726     & 2\%  \\
    \midrule
    Lenses & HSC Strong Lenses  & Legacy Survey  & 758     & 0.1\% \\
    \bottomrule
  \end{tabular}
  \caption{High-confidence query sets used for our retrieval experiments.  
           Citizen-science morphologies (spirals, mergers) come from the Galaxy Zoo–DECaLS catalog \citep{walmsley2022galaxy}, while strong-lens candidates are drawn from previous lens-finding catalog in HSC crossmatched with Legacy Survey \citep{SUGOHI_X}. All images used are from the Legacy Survey in the ensuing results.}
  \label{tab:retrieval_query_stats}
\end{table}

\paragraph{Strong-Lens Retrieval Experiments} For the strong lensing retrieval task, we start by filtering the cross-matched catalog of objects within the Legacy Survey and HSC datasets to approximately reproduce the parent sample used in the HSC strong lensing searches \cite{SUGOHI_X}. Specifically, we impose three additional cuts: (1) objects with photometric redshifts between 0.2 and 1.2, (2) objects with an estimated stellar mass above $5\times 10^{10} M_\odot$, and (3) objects with a star formation rate to stellar mass ratio less than $1 \times 10^{-10}$. In order to identify the strong gravitational lenses within the resulting parent sample, we cross-match with previous lens-finding catalogs \cite{SUGOHI_I,SUGOHI_II,SUGOHI_III,SUGOHI_V,SUGOHI_VI,SUGOHI_VIII,SUGOHI_X,HOLISMOKES_VI,HOLISMOKES_VIII,HOLISMOKES_XIII,rui2019,talbot2021,bina2016,bolton2006,bolton2008,ruff2011,more2014,Faure2008,Jacobs2019,Goullaud2018,Oguri2012}. This yields roughly 700 strong lenses. Most strong lensing catalogs offer a grade for each candidate. Since the criteria for this grading varies between catalogs, we ignore these grades and instead assign a relevance score of $1.0$ to all the strong lensing candidates found within each catalog. All other objects in our parent sample are given a relevance score of $0.0$. For each strong lens Legacy Survey image, we perform the same steps as above, and return the top $k=10$ candidates, after which we compute the $\mathrm{nDCG@}10$ score and average over all queries.

\begin{figure}[ht]
  \centering
  \begin{subfigure}[ht]{0.51\textwidth}
    \centering
    \includegraphics[width=\linewidth]{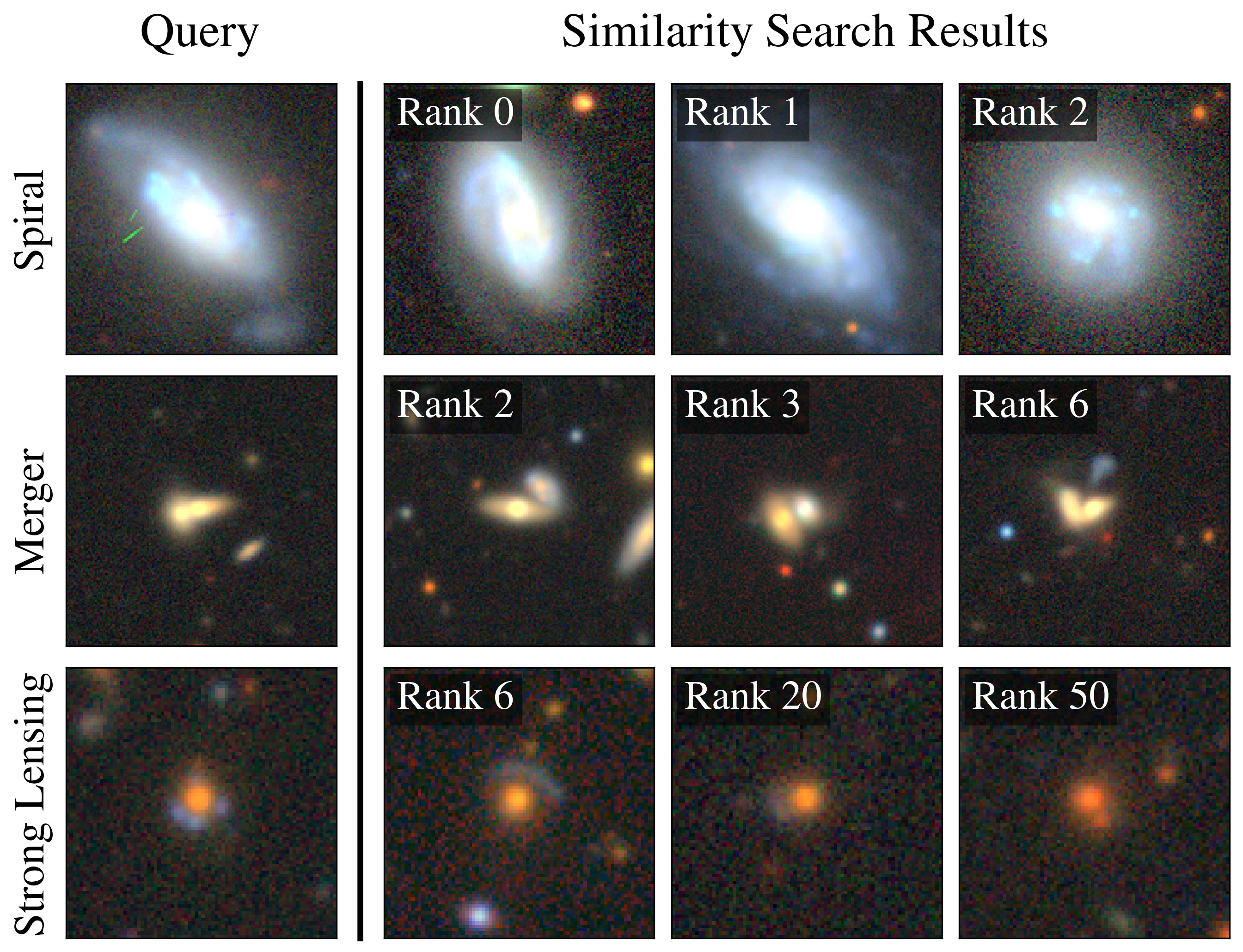}
    \label{fig:retrieval-examples}
  \end{subfigure}%
  \hfill
  \begin{subfigure}[ht]{0.44\textwidth}
    \centering
    \small
    \begin{tabular}{lccc}
        \toprule
                         & \textbf{Spirals} & \textbf{Mergers} & \textbf{Lenses} \\
        \midrule
        AION-1-B         & 0.632 & 0.281 & 0.012 \\
        AION-1-L         & 0.643 & 0.303 & 0.011 \\
        AION-1-XL        & 0.621 & 0.384 & 0.015 \\
        \midrule
        \citet{Parker2024} & 0.602 & 0.248 & 0.006 \\
        \citet{stein2021self} & 0.590 & 0.340 & 0.007 \\
        \citet{oquab2023dinov2} & 0.477 & 0.060 & 0.003 \\
        \bottomrule
      \end{tabular}
    \label{tab:retrieval-scores}
  \end{subfigure}
  \caption{\textbf{Galaxy Image Retrieval} for three astronomical classes of decreasing prevalence; spirals, mergers and strong lenses. \textit{Left}: Example candidates for a given query image. Correct candidates are shown, along with their rank among all retrieved objects sorted by cosine similarity. \textit{Right}: The $\mathrm{nDCG@}10$ score for each of the classes averaged over all queries in the given dataset.}
  \label{fig:retrieval_qual_quant}
\end{figure}

\paragraph{Results} \autoref{fig:retrieval_qual_quant} demonstrates the performance of using AION-1's encoder as an embedding model for rare object retrieval tasks. We include for comparison the results from two state-of-the-art astrophysics foundation models, AstroCLIP \citep{Parker2024} and a self-supervised galaxy image model on Legacy Survey images \citep{stein2021self}, which are executed and evaluated in the same way as AION-1, except for the fact that they both ingest $\{g,r,z\}$ Legacy Survey imaging rather than $\{g,r,i,z\}$ imaging. We also demonstrate the relative performance of a DINOv2 \citep{oquab2023dinov2} vision model's embeddings on these tasks. Qualitatively, the nearest neighbours returned by AION-1-XL are visually convincing for all three classes—even for strong lenses, whose true abundance is below 0.1\%. Quantitatively, AION-1 outperforms every baseline across the board. Taken together, these results demonstrate that AION-1’s latent space is sufficiently discriminative to enable zero-shot retrieval of both common and vanishingly rare phenomena.

\subsection{Emergent Transfer Properties}
\label{sec:transfer}

\subsubsection{Generative Transfer of Multimodal Understanding}

\begin{figure}
  \centering
  \includegraphics[width=0.85\linewidth]{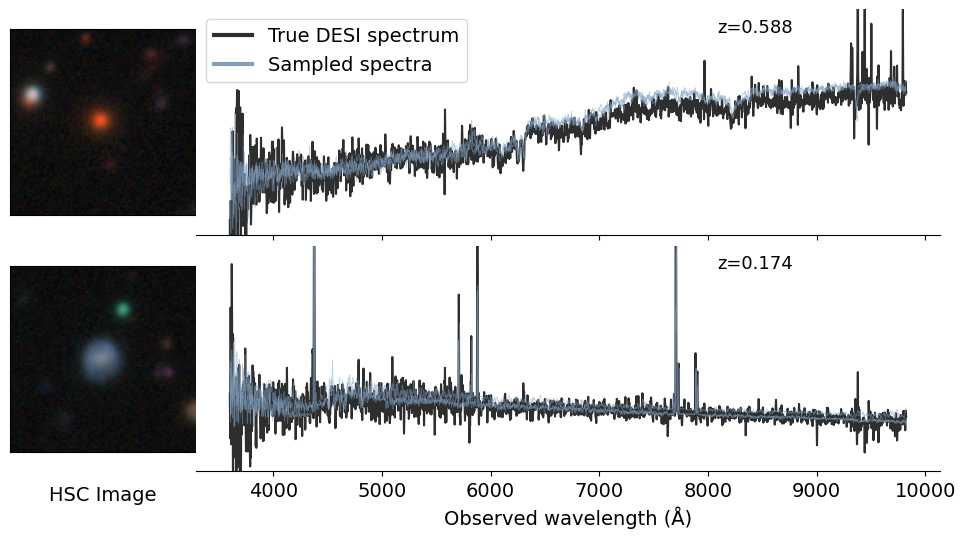}
  \caption{\textbf{Out-of-distribution conditional generation.}  
  DESI spectra sampled from \textsc{AION-1} (blue) conditioned on HSC images (insets), overlaid on the true DESI spectra (black).  
  Even though HSC–DESI pairs were never seen during pre-training, the model reproduces key spectral features, demonstrating emergent transitive understanding.}
  \label{fig:hsc_desi_transfer}
\end{figure}

Our training mixture contains many \emph{pairwise} matches, e.g.\ HSC images $\leftrightarrow$ SDSS spectra and SDSS spectra $\leftrightarrow$ DESI spectra. However, the model is never trained to on the DESI-HSC pairs, and so never learns to produce HSC images from DESI spectra directly. Nevertheless, when we condition use AION-1 to sample a DESI spectrum from an HSC image, the generated spectrum (blue) closely tracks the ground-truth DESI spectrum (black); see \autoref{fig:hsc_desi_transfer}. Both a quiescent (red) galaxy and a star-forming (blue) galaxy are reproduced with realistic absorption and emission features, demonstrating that AION-1 has learned a transitive mapping across modalities. This is likely due to the fact that AION-1 already understands the mapping between HSC and other spectroscopic (SDSS) or imaging (Legacy Survey) surveys, as well as the mapping between those intermediate surveys and the final target survey, DESI.

\subsubsection{Survey-to-survey Transfer in Embedding Space}

Beyond conditional generation, we ask whether AION-1’s frozen image encoder produces survey-invariant representations that let us port knowledge from one telescope to another. Specifically, we train a single linear classifier on Legacy Survey embeddings to predict the ten Galaxy Zoo-10 morphology classes (see \autoref{sec:galaxy-morphology-classification}).  The encoder weights remain fixed; only the 10-way soft-max layer is optimized.  We then apply this exact head—\emph{without any fine-tuning}—to embeddings of Hyper Suprime-Cam (HSC) images. To create the evaluation set we cross-match the HSC wide catalog (see \autoref{sec:data:hsc} with Galaxy Zoo-10 (GZ10) volunteer votes and explicitly remove any targets that overlap with the Legacy Survey-GZ10 training set to prevent test leakage.  The resulting sample contains ${\sim}$1,000 galaxies. As reported in \autoref{tab:transfer_ls_hsc}, the zero-shot classifier attains $84 - 86\%$ accuracy across all AION-1 scales, essentially matching its performance on the native Legacy Survey domain. This robustness holds despite factor-of-$\sim$2 differences in depth, distinct filter sets ($\{g,r,i,z,y\}$ vs.\ $\{g,r,i,z\}$), and a different pixel scale. The result underscores that AION-1 embeddings capture morphology in a way that is largely agnostic to survey-specific imaging characteristics, enabling workflows that recycle scarce labeled data from one survey to bootstrap science in another.

\begin{table}[ht]
  \centering
  \small
  \begin{tabular}{lcc}
    \toprule
                  & \textbf{Legacy Survey (Train)} & \textbf{HSC (Eval.)} \\
    \midrule
    AION-1-B      & 83.95 & 84.15 \\
    AION-1-L      & 87.16 & 85.66 \\
    AION-1-XL     & 86.99 & 85.91 \\
    \bottomrule
  \end{tabular}
  \caption{\textbf{Zero-shot morphology-classification accuracy (\%).}  
           A classifier trained on Legacy Survey images transfers directly to HSC, indicating that AION-1 produces survey-invariant representations.}
  \label{tab:transfer_ls_hsc}
\end{table}

\section{Conclusion}
In this work we have introduced \textbf{AION‑1}, a billion–parameter, omni-modal foundation‐model family for the astronomical sciences. Leveraging a series of tokenizers to homogenise 39 heterogeneous data modalities from five of the largest public surveys, and a multimodal masked-modeling objective to learn their joint distribution, AION‑1 demonstrates—for the first time—the feasibility and utility of training a single, encoder-decoder architecture that:
\begin{itemize}
    \item \textbf{Integrates imaging, spectroscopy and scalar metadata} drawn from instruments that differ widely in resolution, noise properties and wavelength coverage, without bespoke per‑task architecture changes.
    \item \textbf{Achieves or surpasses state‑of‑the‑art performance} on a diverse suite of downstream tasks—including galaxy and stellar property estimation, morphology classification, image segmentation and spectral super‑resolution—using only lightweight probes or small task‑specific heads.
    \item \textbf{Excels in the low‑data regime}, maintaining high $R^{2}$ and classification accuracy with two to three orders of magnitude fewer labels than fully supervised baselines, a critical capability when high‑quality annotations are scarce or expensive.
    \item \textbf{Enables zero‑shot semantic retrieval}, discovering rare objects such as strong gravitational lenses with better scores than previous state-of-the-art astronomical foundation models.
    \item \textbf{Enables transfer learning} across surveys by producing survey-invariant representations allowing us to port downstream tasks from one telescope to another.
\end{itemize}
Ultimately, AION‑1 offers the community a new backbone that collapses traditional, siloed pipelines into a single model. Astronomers can now fuse heterogeneous observations, prototype new analyses and mine extreme outliers with only modest computational resources. By releasing all code, tokenisers, pretrained weights and evaluation suites under an open‑source licence, we hope to accelerate the adoption of foundation‑model approaches across current and forthcoming surveys.

\subsection{Limitations and future work}
\label{subsec:limitations}
Despite its versatility, AION‑1 inherits several caveats. We discuss these in detail below.

\paragraph{Architectural Limitations} There are two main architectural limitations to the AION-1 model: (1) \emph{Discretisation limits}: The use of quantization/discretization during the tokenization process intrinisically limits the information content captured, which may impact downstream analyses; and (2) \emph{Naïve embedding aggregation}: the mean-pool strategy used for retrieval is a pragmatic first pass but leaves performance on the table; contrastive post-training or other more intelligent approaches could furnish richer, task-aware representations.

\paragraph{Selection Functions \& Representativeness} AION-1 inherits the selection functions of its pre-training data: magnitude and quality cuts (e.g., LS magnitude thresholds, HSC ``full-depth full-colour'' + flags), Gaia XP availability, DESI EDR SV3 filters, sky-footprint choices (e.g., SGC-only), and our reciprocal cross-match. These selection functions may influence downstream predictions in the absence of recalibration  \citep{kumar2022finetuningdistortpretrainedfeatures, gallegos2024biasfairnesslargelanguage, thaler2024farbiasgo}. For this reason, we primarily propose using AION-1 as an embedding model to extract relevant features from a given data source, on top of which we train a simple projection head which is effectively re-calibrated during task-specific adaptation on a specific downstream dataset that defines both a task and a selection function for that task. Nonetheless, future work on exploring selection functions in pretraining would be an exciting direction for the astronomy community. 

\paragraph{Generative Capabilities} Our pre-training uses multimodal masked modeling. While this is effective and simple, it does not furnish a principled joint sampler for modalities with many output tokens (images, spectra, segmentation maps). In practice, samples can look convincing and even score well on task metrics, yet still be mis-calibrated and/or under-correlated across tokens. This is not a new problem—masked-model decoders and, more broadly, many ML generative methods face calibration gaps in multi-token settings, including in astronomy. Accordingly, we mainly suggest that the astronomy community recalibrate generative tasks using AION-1 embeddings instead of performing posterior draws over multi-token outputs so that downstream samples are better calibrated and preserve the correct token-to-token correlations prior to scientific use. 

For future work, there exist alternatives to masked modeling. For example, one can replace masked decoding with models that define a coherent joint distribution over output tokens—e.g., (i) autoregressive token models (LM/decoder‑only) with proper likelihoods, or (ii) diffusion models over continuous or tokenized representations conditioned on observed modalities. Both choices typically yield better‑behaved samplers and correlations across tokens. However, they still encode a pre‑training prior that may not match a new survey or selection; principled adaptation is then necessary; see, for instance, \cite{rozet2024learning} for an expectation-maximization approach to refit a diffusion prior from incomplete or shifted observations.

\subsection{Broader scientific impact}
Although developed for astronomy, the AION‑1 recipe—data‑driven tokenisation, scale‑appropriate masked modelling and modality‑aware provenance embeddings—addresses challenges endemic to many experimental sciences: heterogeneity, noise and instrument‑specific idiosyncrasies. We therefore envisage direct extensions to adjacent domains in which multi‑instrument data proliferation similarly outpaces bespoke modeling.

\section{Contributions}
The author contributions are summarized below. In each category, authors are listed alphabetically.

\begin{itemize}
    \item \textbf{Project Leads}: Francois Lanusse and Liam Parker
    \item \textbf{Data Team}: Micah Bowles, Tom Hehir, Lucas Meyer, Francois Lanusse, Liam Parker, Jeff Shen, Helen Qu, Sebastian Wagner-Carena
    \item \textbf{Tokenization Team}:
    \begin{itemize}
        \item \textbf{Image}: Francois Lanusse, Liam Parker
        \item \textbf{Spectrum}: Francois Lanusse, Jeff Shen
        \item \textbf{Scalar}: Jeff Shen, Sebastian Wagner-Carena
        \item \textbf{Catalog}: Ollie Liu
        \item \textbf{Segmentation maps}: Micah Bowles, Tom Hehir
    \end{itemize}
    \item \textbf{Pretraining Team}: Francois Lanusse, Siavash Golkar, Liam Parker, Leopoldo Sarra
    \item \textbf{Downstream Evaluation}: Micah Bowles, Tom Hehir, Francois Lanusse, Ollie Liu, Lucas Meyer, Liam Parker, Jeff Shen, Helen Qu, Sebastian Wagner-Carena
    \item \textbf{Computing and Optimization}: Hatim Bourfoune, Nathan Cassereau, Pierre Cornette, Geraud Krawezik, Lucas Meyer, Ruben Ohana
    \item \textbf{Manuscript Writing}: Micah Bowles, Tom Hehir, Francois Lanusse, Ollie Liu, Lucas Meyer, Liam Parker, Jeff Shen, Helen Qu, Sebastian Wagner-Carena
    \item \textbf{Advisory}: Alberto Bietti, Kyunghyun Cho, Miles Cranmer, Shirley Ho
\end{itemize}

\section{Acknowledgments}

We would like to acknowledge the support of the Simons Foundation and of Schmidt Sciences. This project was provided with computer and storage resources by GENCI at IDRIS thanks to the grant 2024-GC011015468 on the supercomputer Jean Zay's H100 partition. Additionally, some of the computations in this work were run at facilities supported by the Scientific Computing Core at the Flatiron Institute, a division of the Simons Foundation. Liam Parker also acknowledges support from the National Science Foundation Graduate Research Fellowship Program. Jeff Shen is supported by the Natural Sciences and Engineering Research Council of Canada (NSERC), funding reference number 587652. We would like to thank Andrew Engel, Marc Huertas-Company, Stephanie Juneau, Andy Morgan, and Mike Smith for their valuable feedback during beta testing and Sophie Barstein for her help with writing the corresponding blog post.

\subsection{Data}
\subsubsection{Legacy Survey}
The Legacy Surveys consist of three individual and complementary projects: the Dark Energy Camera Legacy Survey (DECaLS; Proposal ID \#2014B-0404; PIs: David Schlegel and Arjun Dey), the Beijing-Arizona Sky Survey (BASS; NOAO Prop. ID \#2015A-0801; PIs: Zhou Xu and Xiaohui Fan), and the Mayall z-band Legacy Survey (MzLS; Prop. ID \#2016A-0453; PI: Arjun Dey). DECaLS, BASS and MzLS together include data obtained, respectively, at the Blanco telescope, Cerro Tololo Inter-American Observatory, NSF’s NOIRLab; the Bok telescope, Steward Observatory, University of Arizona; and the Mayall telescope, Kitt Peak National Observatory, NOIRLab. Pipeline processing and analyses of the data were supported by NOIRLab and the Lawrence Berkeley National Laboratory (LBNL). The Legacy Surveys project is honored to be permitted to conduct astronomical research on Iolkam Du’ag (Kitt Peak), a mountain with particular significance to the Tohono O’odham Nation.

NOIRLab is operated by the Association of Universities for Research in Astronomy (AURA) under a cooperative agreement with the National Science Foundation. LBNL is managed by the Regents of the University of California under contract to the U.S. Department of Energy.

This project used data obtained with the Dark Energy Camera (DECam), which was constructed by the Dark Energy Survey (DES) collaboration. Funding for the DES Projects has been provided by the U.S. Department of Energy, the U.S. National Science Foundation, the Ministry of Science and Education of Spain, the Science and Technology Facilities Council of the United Kingdom, the Higher Education Funding Council for England, the National Center for Supercomputing Applications at the University of Illinois at Urbana-Champaign, the Kavli Institute of Cosmological Physics at the University of Chicago, Center for Cosmology and Astro-Particle Physics at the Ohio State University, the Mitchell Institute for Fundamental Physics and Astronomy at Texas A\&M University, Financiadora de Estudos e Projetos, Fundacao Carlos Chagas Filho de Amparo, Financiadora de Estudos e Projetos, Fundacao Carlos Chagas Filho de Amparo a Pesquisa do Estado do Rio de Janeiro, Conselho Nacional de Desenvolvimento Cientifico e Tecnologico and the Ministerio da Ciencia, Tecnologia e Inovacao, the Deutsche Forschungsgemeinschaft and the Collaborating Institutions in the Dark Energy Survey. The Collaborating Institutions are Argonne National Laboratory, the University of California at Santa Cruz, the University of Cambridge, Centro de Investigaciones Energeticas, Medioambientales y Tecnologicas-Madrid, the University of Chicago, University College London, the DES-Brazil Consortium, the University of Edinburgh, the Eidgenossische Technische Hochschule (ETH) Zurich, Fermi National Accelerator Laboratory, the University of Illinois at Urbana-Champaign, the Institut de Ciencies de l’Espai (IEEC/CSIC), the Institut de Fisica d’Altes Energies, Lawrence Berkeley National Laboratory, the Ludwig Maximilians Universitat Munchen and the associated Excellence Cluster Universe, the University of Michigan, NSF’s NOIRLab, the University of Nottingham, the Ohio State University, the University of Pennsylvania, the University of Portsmouth, SLAC National Accelerator Laboratory, Stanford University, the University of Sussex, and Texas A\&M University.

BASS is a key project of the Telescope Access Program (TAP), which has been funded by the National Astronomical Observatories of China, the Chinese Academy of Sciences (the Strategic Priority Research Program “The Emergence of Cosmological Structures” Grant \# XDB09000000), and the Special Fund for Astronomy from the Ministry of Finance. The BASS is also supported by the External Cooperation Program of Chinese Academy of Sciences (Grant \# 114A11KYSB20160057), and Chinese National Natural Science Foundation (Grant \# 12120101003, \# 11433005).

The Legacy Survey team makes use of data products from the Near-Earth Object Wide-field Infrared Survey Explorer (NEOWISE), which is a project of the Jet Propulsion Laboratory/California Institute of Technology. NEOWISE is funded by the National Aeronautics and Space Administration.

The Legacy Surveys imaging of the DESI footprint is supported by the Director, Office of Science, Office of High Energy Physics of the U.S. Department of Energy under Contract No. DE-AC02-05CH1123, by the National Energy Research Scientific Computing Center, a DOE Office of Science User Facility under the same contract; and by the U.S. National Science Foundation, Division of Astronomical Sciences under Contract No. AST-0950945 to NOAO.

\subsubsection{Hyper Suprime-Cam}
The Hyper Suprime-Cam (HSC) collaboration includes the astronomical communities of Japan and Taiwan, and Princeton University. The HSC instrumentation and software were developed by the National Astronomical Observatory of Japan (NAOJ), the Kavli Institute for the Physics and Mathematics of the Universe (Kavli IPMU), the University of Tokyo, the High Energy Accelerator Research Organization (KEK), the Academia Sinica Institute for Astronomy and Astrophysics in Taiwan (ASIAA), and Princeton University. Funding was contributed by the FIRST program from Japanese Cabinet Office, the Ministry of Education, Culture, Sports, Science and Technology (MEXT), the Japan Society for the Promotion of Science (JSPS), Japan Science and Technology Agency (JST), the Toray Science Foundation, NAOJ, Kavli IPMU, KEK, ASIAA, and Princeton University. 

This paper makes use of software developed for the Large Synoptic Survey Telescope. We thank the LSST Project for making their code available as free software at  http://dm.lsst.org

The Pan-STARRS1 Surveys (PS1) have been made possible through contributions of the Institute for Astronomy, the University of Hawaii, the Pan-STARRS Project Office, the Max-Planck Society and its participating institutes, the Max Planck Institute for Astronomy, Heidelberg and the Max Planck Institute for Extraterrestrial Physics, Garching, The Johns Hopkins University, Durham University, the University of Edinburgh, Queen’s University Belfast, the Harvard-Smithsonian Center for Astrophysics, the Las Cumbres Observatory Global Telescope Network Incorporated, the National Central University of Taiwan, the Space Telescope Science Institute, the National Aeronautics and Space Administration under Grant No. NNX08AR22G issued through the Planetary Science Division of the NASA Science Mission Directorate, the National Science Foundation under Grant No. AST-1238877, the University of Maryland, and Eotvos Lorand University (ELTE) and the Los Alamos National Laboratory.

\subsubsection{Dark Energy Spectroscopic Instrument}
This research used data obtained with the Dark Energy Spectroscopic Instrument (DESI).
DESI construction and operations is managed by the Lawrence Berkeley National Laboratory.
This material is based upon work supported by the U.S. Department of Energy, Office of
Science, Office of High-Energy Physics, under Contract No. DE–AC02–05CH11231, and
by the National Energy Research Scientific Computing Center, a DOE Office of Science
User Facility under the same contract. Additional support for DESI was provided by the
U.S. National Science Foundation (NSF), Division of Astronomical Sciences under Contract
No. AST-0950945 to the NSF’s National Optical-Infrared Astronomy Research Laboratory;
the Science and Technology Facilities Council of the United Kingdom; the Gordon and
Betty Moore Foundation; the Heising-Simons Foundation; the French Alternative Energies
and Atomic Energy Commission (CEA); the National Council of Science and Technology
of Mexico (CONACYT); the Ministry of Science and Innovation of Spain (MICINN), and
by the DESI Member Institutions: www.desi.lbl.gov/collaborating-institutions. The DESI
collaboration is honored to be permitted to conduct scientific research on Iolkam Du’ag (Kitt
Peak), a mountain with particular significance to the Tohono O’odham Nation. Any opinions,
findings, and conclusions or recommendations expressed in this material are those of the
author(s) and do not necessarily reflect the views of the U.S. National Science Foundation,
the U.S. Department of Energy, or any of the listed funding agencies.

\subsubsection{Sloan Digital Sky Survey}
Funding for the Sloan Digital Sky Survey V has been provided by the Alfred P. Sloan Foundation, the Heising-Simons Foundation, the National Science Foundation, and the Participating Institutions. SDSS acknowledges support and resources from the Center for High-Performance Computing at the University of Utah. SDSS telescopes are located at Apache Point Observatory, funded by the Astrophysical Research Consortium and operated by New Mexico State University, and at Las Campanas Observatory, operated by the Carnegie Institution for Science. The SDSS web site is \url{www.sdss.org}.

SDSS is managed by the Astrophysical Research Consortium for the Participating Institutions of the SDSS Collaboration, including Caltech, The Carnegie Institution for Science, Chilean National Time Allocation Committee (CNTAC) ratified researchers, The Flatiron Institute, the Gotham Participation Group, Harvard University, Heidelberg University, The Johns Hopkins University, L'Ecole polytechnique f\'{e}d\'{e}rale de Lausanne (EPFL), Leibniz-Institut f\"{u}r Astrophysik Potsdam (AIP), Max-Planck-Institut f\"{u}r Astronomie (MPIA Heidelberg), Max-Planck-Institut f\"{u}r Extraterrestrische Physik (MPE), Nanjing University, National Astronomical Observatories of China (NAOC), New Mexico State University, The Ohio State University, Pennsylvania State University, Smithsonian Astrophysical Observatory, Space Telescope Science Institute (STScI), the Stellar Astrophysics Participation Group, Universidad Nacional Aut\'{o}noma de M\'{e}xico, University of Arizona, University of Colorado Boulder, University of Illinois at Urbana-Champaign, University of Toronto, University of Utah, University of Virginia, Yale University, and Yunnan University.

\subsubsection{Gaia}
This work has made use of data from the European Space Agency (ESA) mission {\it Gaia} (\url{https://www.cosmos.esa.int/gaia}), processed by the {\it Gaia} Data Processing and Analysis Consortium (DPAC, \url{https://www.cosmos.esa.int/web/gaia/dpac/consortium}). Funding for the DPAC has been provided by national institutions, in particular the institutions participating in the {\it Gaia} Multilateral Agreement. The Gaia data are open and free to use, provided credit is given to `ESA/Gaia/DPAC'. If you use Gaia DR3 data in your research, please acknowledge it as above.

\bibliographystyle{assets/plainnat}
\bibliography{mmoma.bib}

\end{document}